\documentclass[11pt,letterpaper]{article}
\usepackage{graphicx} 
\usepackage{parskip}
\usepackage[symbol]{footmisc}
\usepackage[dvipsnames]{xcolor}
\usepackage{amsmath, amssymb, amsfonts, hyperref, amsthm, listings, tcolorbox, bbm, xifthen, mathtools, titlesec, graphicx, pdfpages}
\usepackage[linesnumbered,lined,boxed,commentsnumbered, ruled]{algorithm2e}
\usepackage[shortlabels]{enumitem}
\usepackage{hyperref}
\hypersetup{
    colorlinks=true,
    linkcolor=blue,
    filecolor=magenta,
    citecolor=magenta,
    urlcolor=cyan,
}
\usepackage[margin=1in]{geometry}
\usepackage[backend=biber, style=alphabetic]{biblatex}
\newcommand{\poly}{\mathrm{poly}}

\newtheorem{theorem}{Theorem}[section]
\newtheorem{lemma}[theorem]{Lemma}

\newtheorem{corollary}[theorem]{Corollary}
\newtheorem{claim}[theorem]{Claim}

\newtheorem{optimization}[theorem]{Optimization}
\theoremstyle{definition}
\newtheorem{definition}[theorem]{Definition}
\newtheorem{example}[theorem]{Example}
\newtheorem{remark}{Remark}

\newcommand{\ceil}[1]{\left\lceil #1 \right\rceil}

\renewcommand{\subset}{\subseteq}
\DeclareMathOperator*{\E}{\ensuremath{\mathbb{E}}}
\DeclareMathOperator{\agree}{agree}
\DeclareMathOperator{\Agree}{Agree}
\DeclareMathOperator{\1}{\ensuremath{\mathbf{1}}}
\DeclareMathOperator{\N}{\mathbb{N}}

\title{An Improved Johnson Bound for List-Decoding with Hints}
\author{Swastik Kopparty \thanks{Department of Mathematics and Department of Computer Science, University of Toronto. Email: \texttt{swastik.kopparty@utoronto.ca}.   Research partially supported
by an NSERC Discovery Grant.} \and Harry Sha \thanks{Department of Computer Science, University of Toronto. Email: \texttt{shaharry@cs.toronto.edu}.   Research partially supported
by an NSERC Discovery Grant.} }
\date{}
\begin{document}

\maketitle

\begin{abstract}
This paper studies when list-decodability guarantees for error-correcting codes of good distance can go beyond the classical Johnson radius and gives a corresponding efficient list-decoding algorithm for Reed-Solomon codes up to this new radius. The Johnson bound shows that any error-correcting code with distance $\delta$ is automatically list-decodable with small list size up to a certain radius $J(\delta)$ around any received word $r$. We show that if the received word $r$ is known to have some codewords that are somewhat close to it (we call these ``hints"), then $C$ is list-decodable with small list size up to an even larger radius around $r$.

We determine the optimal list-decoding radius in this setting for single and multiple hints over both large and small alphabets and show that this radius is the optimum of a natural convex program. Remarkably, even a single hint that is $(J(\delta)-\epsilon)$-close to $r$ improves the list-decoding radius around $r$ to $J(\delta)+\epsilon'$.

Finally, we give an efficient algorithm, based on the weighted Guruswami-Sudan decoder, that list-decodes Reed-Solomon codes up to this optimal radius.
\end{abstract}

\section{Introduction}

This paper studies when list-decodability guarantees for error-correcting codes of good distance can go beyond the classical Johnson radius and gives a corresponding efficient list-decoding algorithm for Reed-Solomon codes up to this new radius. Very briefly,
the Johnson bound shows that for any code and any received word $r$, the ball around $r$
of a certain radius contains very few codewords. We show that if the received word $r$
is known to have some codewords that are somewhat close to it, then a ball of an even larger radius around $r$ contains very few codewords. We give a sharp characterization of this radius and use our understanding to give {\em efficient} algorithms for list-decoding Reed-Solomon codes up to this larger radius\footnote{This paper was written before the algorithmic breakthroughs
on list-decoding Reed-Solomon codes over prime fields~\cite{BCPZZ}, see also \cite{HKS}. We remark that our algorithmic results for Reed-Solomon codes are in the regime of {\em constant} list size bounds for decoding beyond the Johnson radius around certain centers, while \cite{BCPZZ} gives a polynomial list size bound around all centers.}

To describe things in more detail, we quickly introduce some notation. For an alphabet $\Sigma$ and large $n$, we work with the metric space $\Sigma^n$ equipped with the relative Hamming distance, where $\Delta(x,y)$ is the fraction of coordinates $i \in [n]$ for which $x_i \neq y_i$.
An error-correcting code of (relative) distance $\delta$ over the alphabet $\Sigma$ is a subset
$C \subseteq \Sigma^n$ such that any two elements of $C$ have distance at least $\delta$. By the triangle inequality, such a code immediately satisfies
unique decodability up to radius $\frac{\delta}{2}$: for any string $r \in \Sigma^n$, there can be at most one element of $C$ that is less than $\delta/2$-close to $r$.
This lets us use $C$ to correct $\delta/2$ errors.

To correct even more errors, we have to relax to the notion of {\em list-decoding}, introduced by Elias in~\cite{Elias}.
Given a received word $r$, the goal of list-decoding is to find all codewords within a given distance $\rho$ of $r$ while ensuring that the number of such codewords is small. The Johnson bound~\cite{johnsonNewUpperBound1962} gives a generic nontrivial combinatorial list-decodability result: for any code $C$ of distance $\delta$, the number of codewords within radius $\rho = 1 - \sqrt{1-\delta}$ of any received word is at most polynomial in the code length. Remarkably, when $\delta$ is a constant sufficiently close to 1, the radius $\rho$ can get arbitrarily close to $1$!

In this generality, the bound is tight: if $\rho$ exceeds this threshold by any constant, then there are codes with exponentially many codewords within radius $\rho$ of some received word. The codes witnessing this are, as far as we know, nonlinear. For linear codes, things are much trickier, but it is still known that there are codes and received words with superpolynomially many codewords at this radius~\cite{guruswami-conditional,guruswamiUnconditionalProofTightness2002}.

Progress on the algorithmic side of list-decoding began with the classical univariate polynomial evaluation codes, Reed-Solomon codes. The deep and fundamental Guruswami-Sudan algorithm~\cite{Sudan,GuruswamiSudan} showed that for Reed-Solomon codes,
it is possible to efficiently list-decode (i.e., find all the $\rho$-close codewords) up to the
Johnson radius $\rho = 1 - \sqrt{1-\delta}$. This result (and the underlying algorithm) had a profound impact on coding theory, pseudorandomness, and complexity theory in the following decades, leading to state-of-the-art constructions of list-decodable codes~\cite{ParvareshVardy,GuruswamiRudra,KoppartyRonZewiSarafWootters,Srivastava,ChenZhang}, extractors~\cite{ShaltielUmans,GUV,DKSS,TaShmaUmans}, and pseudorandom generators~\cite{STV,ShaltielUmans,Umans}, among others.

Despite all this impact, the list-decodability of Reed-Solomon codes itself remains shrouded in mystery. It is known that some Reed-Solomon codes do need list sizes polynomially large in $n$ when decoding slightly beyond the Johnson radius~\cite{BenSassonKoppartyRadhakrishnan}.
It is also known that Reed-Solomon codes with random evaluation points are list-decodable up to any radius $\rho < \delta$ with constant list size~\cite{BrakensiekGopiMakam,GuoZhang,GuruswamiAlrabiahLi}. It remains a tantalizing possibility that every Reed-Solomon code can be list-decoded up to any radius $\rho < \delta$ with $\poly(n)$ list size, and furthermore, that this list can be found efficiently\footnote{We now know, from~\cite{BCPZZ} (see also~\cite{HKS}), that this is true over prime fields.}.

Given this state of affairs, it is of great interest to find new methods that {\em prove} or {\em certify} that for a given $r \in \Sigma^n$ and radius $\rho$, the number of codewords within distance $\rho$ of $r$ is small. There are two known methods for this. The first is the Johnson bound, which applies uniformly to all $r$ and has $\rho$ equal to the Johnson radius $J(\delta)$. The second is the triangle inequality: if we have a codeword
$h \in C$ with $\Delta(r, h) < \delta-\rho$, then there are no other codewords within distance $\rho$ of $r$.

In this paper, we give a new method (that applies to all codes) to {\em prove} that the list of codewords within a radius $\rho \geq J(\delta)$ of a given received word $r$ is small. This method is based on codewords $h \in C$ that are somewhat close to $r$. We call such codewords {\em hints}. The more hints we know about and the closer they are to $r$, the larger the radius $\rho$ will be. We then show tightness of this improved radius $\rho$ and give an efficient algorithm to list-decode Reed-Solomon codes up to this radius $\rho$.

We summarize our main results below:
\begin{itemize}
    \item For general codes $C$, for every $\epsilon > 0$, there is an $\epsilon' > 0$ such that if there exists even one $h \in C$ with
	    $\Delta(h,r) < J(\delta) - \epsilon$, then list-decoding around $r$ to radius $J(\delta)+\epsilon'$ produces only constant-size lists.
    \item For general codes $C$, given a collection of $h_1, \ldots, h_m \in C$ with
	    $\Delta(h_i, r) < J(\delta)$, we show how to compute a radius $\rho$ depending on the relative geometry of the $h_i$ and $r$ so that list-decoding $C$ to radius $\rho$ around $r$ produces only constant-size lists.
	    \item Conversely, we show that the radius $\rho$ computed above is tight in general: for any desired relative geometry that is realizable, there are codes $C$, received words $r$, and codewords $h_1, \ldots, h_m$ realizing this relative geometry for which list decoding beyond the radius $\rho$ produces lists of size exponential in $n$.
	    \item The above results generalize the large-alphabet (a.k.a. alphabet-independent) Johnson bound. We also prove small-alphabet versions of these results, generalizing the small-alphabet Johnson bound.
    \item Finall, we give efficient algorithms for list-decoding Reed-Solomon codes up to this radius $\rho$ (which, because of the combinatorial results above, results in a constant-size list).
\end{itemize}

The closest point of comparison is the {\em average radius} Johnson bound.
This is a folklore generalization of the Johnson bound that states that
for any code $C$ with distance $\delta$, for any $r \in \Sigma^n$, and for any $L$ codewords $c_1, \ldots, c_L \in C$, their average distance to $r$ must be at least the Johnson radius:
$$ \frac{1}{L}\sum_{i=1}^L \Delta(c_i, r)  \geq J\left(\delta - \frac{\delta}{L} \right)= J(\delta) - o_L(1).$$
It is completely consistent with the average radius Johnson bound that for some constant $\epsilon>0$, one
codeword has distance $J(\delta)-\epsilon$ to $r$, while there are exponentially many
codewords with distance exactly $J(\delta)$ to $r$.

In contrast, the simplest instantiation of our improved list-decoding bound rules this out.
Even one codeword with distance $J(\delta)-\epsilon$ to $r$ forces the number of
codewords within distance $J(\delta) + \epsilon'$ of $r$ (where $\epsilon'>0$ is a constant depending on $\epsilon$ and $\delta$) to be bounded by a constant.

For this simple setting, our algorithmic results for Reed-Solomon codes have the following curious consequence. Suppose we are given a received word $r$ for a Reed-Solomon code with
distance $\delta = 0.99$. The Johnson radius for this setting is $$J(\delta) = 1 - \sqrt{1-0.99} = 0.9,$$
and thus the Guruswami-Sudan list-decoder can find any codeword within this radius of $r$.
Now suppose we additionally know that there is some codeword $h$ of the Reed-Solomon code with
$\Delta(h, r) = 0.85$ -- so $h$ only agrees with $r$ on $15\%$ of the coordinates. This fact enables us to efficiently list-decode around $r$ to an even larger radius, $\rho = 0.901$!

We would like to mention one final motivation for results of this type. These results give new and nontrivial information about the set of realizable Hamming distances between codewords of codes $C \subseteq \Sigma^n$ and general elements of $\Sigma^n$. These kinds of questions arise naturally in (1) understanding basic questions about rates, distances, and distance distributions of error-correcting codes (as in the paper of Kalai and Linial~\cite{KalaiLinial-Distance-Distribution}), and (2) decoding concatenated codes using soft information (as in the paper of Guruswami and Sudan~\cite{GS-concatenated-soft-information}). We believe our results and methods could be helpful in the study of these topics.

\subsection{Results}

It is most convenient to state our results in terms of agreements rather than distance.
For $x,y \in \Sigma^n$, we define $\agree(x,y)$ to be the fraction of coordinates $i \in [n]$ such that $x_i = y_i$. We then work with the {\em maximum agreement} $\alpha$ of a code $C$
rather than the minimum distance $\delta$ ($\alpha = 1-\delta$).

We say that a code $C \subseteq \Sigma^n$ is {\em $(\beta, L)$-list-decodable at $r \in \Sigma^n$} if the number of codewords $c \in C$ with $\agree(c,r) \geq \beta$ is at most $L$.

In these terms, the Johnson bound says that any code $C \subseteq \Sigma^n$ with maximum agreement $\alpha$ is $(\sqrt{\alpha}, O(n))$-list-decodable at {\em every} $r \in \Sigma^n$. The main object of this paper is to understand how much the list-decodability of a code $C$ at a certain $r$ improves (over the Johnson bound) if we know that there are some ``hint" codewords $h_1, \ldots, h_m \in C$ that are somewhat close to $r$.


The relationship between the hints and the received word is captured by the {\em agreement profile} $\eta$. For a single hint, $\eta$ is simply the agreement between the hint and the received word; for multiple hints, $\eta$ encodes the full pattern of agreements among the hints and the received word.

Our main (and most general) result is a tight characterization of the agreement $\beta$ for which we are guaranteed list-decodability with small list size at $r$: it is expressed as the optimum of a natural convex program whose constraints are determined by $\eta$ and code parameters such as the maximum agreement and the alphabet size.

\textbf{Theorem 1. } (Theorem~\ref{thm:large-alphabet-johnson-multi}, restated)
{\em
    Let $C \subset \Sigma^n$ be a code of maximum agreement $\alpha$ (minimum distance $1-\alpha$), where $q \coloneqq |\Sigma|$, and assume $\alpha > 1/q$.
    Let $r \in \Sigma^n$ be a received word, and let $h_1,\ldots,h_m \in C$ be hints, with agreement profile $\eta$.

    Let $\beta^*$ be the optimum of the convex program Optimization~\ref{opt:multi}.

    Then for any $\beta > \beta^*$, there is a constant $L = L(\beta, \beta^*)$ such that
    $C$ is $(\beta, L)$-list-decodable at $r$.

    Conversely, for any $\beta < \beta^*$ and any realizable profile $\eta$, there exists a code $C \subseteq \Sigma^n$, a
	    received word $r \in \Sigma^n$, and hints $h_1, \ldots, h_m \in C$ with
    agreement profile $\eta$ such that
    $C$ is {\bf not} $(\beta, \exp(\Omega(n)))$-list-decodable at $r$.}

Note that when the number of hints $m$ is constant,
Optimization~\ref{opt:multi} is a convex program in a constant number of variables with a constant number of constraints.

Optimization~\ref{opt:multi} can be complicated in general, but simplifies considerably in natural special cases.
We study several natural special cases: a single hint over a large alphabet (Optimization~\ref{opt:johnson-large-alphabet-single-hint}), a single hint over a small alphabet (Optimization~\ref{opt:johnson-small-alphabet-single-hint}), and multiple hints over a large alphabet (Optimization~\ref{opt:multi-hint-large-alphabet}).
For a single hint over a large alphabet, we solve the optimization analytically.
The resulting threshold $\beta^*$ is strictly less than the Johnson agreement $\sqrt{\alpha}$ whenever the hint's agreement with the received word exceeds the Johnson agreement $\sqrt{\alpha}$.

\textbf{Theorem 2. } (Theorem~\ref{thm:solve-single-opt}, restated)
    { \em Let $C \subset \Sigma^n$ be a code with maximum agreement $\alpha$ over a large alphabet, let $r \in \Sigma^n$ be a received word, and let $h \in C$ be a hint with $\agree(r, h) = \eta \in [0,1]$.

    Define $\beta^*(\alpha, \eta)$ by
    \[
        \beta^* = \begin{cases}
            \sqrt{\alpha}, & \eta \le \sqrt{\alpha}, \\
            \alpha + \sqrt{(1-\eta)\left(\alpha - \frac{\alpha^2}{\eta}\right)}, & \sqrt{\alpha} < \eta \le \eta_{\mathrm{crit}}, \\
            \alpha + 1 - \eta, & \eta > \eta_{\mathrm{crit}},
        \end{cases}
    \]
    where $\eta_{\mathrm{crit}} = \frac{(1-\alpha) + \sqrt{1-2\alpha+5\alpha^2}}{2}$.
	  (In particular, observe that $\beta^* < \sqrt{\alpha}$ whenever $\eta > \sqrt{\alpha}$.)

    Then for any $\beta > \beta^*$, there is a constant $L = L(\beta, \beta^*)$ such that $C$ is $(\beta, L)$-list-decodable at $r$.

   Conversely, for any $\beta < \beta^*$ and $\eta \in [0,1]$, there exists a code $C \subseteq \Sigma^n$, a
	    received word $r \in \Sigma^n$, and a hint $h\in C$ with
    $\agree(r,h) = \eta$ such that
    $C$ is {\bf not} $(\beta, \exp(\Omega(n)))$-list-decodable at $r$.}

These combinatorial results were stated above for list-decodability. We remark that they also hold for the slightly stronger notion of average-list-decodability (see Definition~\ref{def:average-list-decodability}).

Finally, we show that these combinatorial bounds can be made algorithmic for Reed-Solomon codes, using the weighted Guruswami-Sudan list decoder.

\textbf{Theorem 3. } (Theorem~\ref{thm:decoding-multi-hint}, restated)
{\em
	    Suppose $C$ is a Reed-Solomon code; let $h_1,\ldots,h_m \in C$ be hints, and let $r$ be a received word.
	    Let $\beta^*$ be the list-decoding threshold from Theorem 1.
    Then for any constant $\beta > \beta^*$, there exists a polynomial-time algorithm that finds all codewords $c \in C$ with $\agree(c,r) \ge \beta$.
}

\paragraph{Remark:}
Chronologically, we first discovered our algorithmic result for Reed-Solomon codes with one hint. Specifically, we asked whether it was possible to use the fact that there is one codeword close to our given received word to improve the guarantees of the Guruswami-Sudan list-decoding algorithm. After we managed to answer this,
further investigation led us to discover that the underlying improved combinatorial bound actually holds for all error-correcting codes.
The initial modification to the Guruswami--Sudan list-decoding algorithm that utilized a nearby codeword did not use weights and instead multiplied the interpolant by the nearby polynomial itself. We discuss this approach in Appendix~\ref{sec:alternate-decoding}.

\subsection{Proof Overview}

We give a high-level overview of the three parts of our argument: the upper bound on list size, the matching lower bound, and the efficient decoding algorithm for Reed-Solomon codes.

\subsubsection{Upper bounds on list size}

We start with the single-hint, large-alphabet case, which contains the main ideas in their simplest form, and then explain how the argument extends to the general setting.

\paragraph{Single hint, large alphabet.}
Let $L \subset C \setminus \{h\}$ be a list of codewords with agreement at least $\beta$ with the received word $r$. Let $p_{i,\sigma} \coloneqq \Pr_{c \sim L}[c_i = \sigma]$ be the fraction of list elements that have symbol $\sigma$ at index $i$.

Since all distinct codewords have pairwise agreement at most $\alpha$,
the expected pairwise agreement of two random elements of $L$ is at most
$\alpha + \frac{1}{|L|}$.

On the other hand, the expected pairwise agreement of two random elements of $L$
equals
$$   \frac{1}{n}\sum_{i,\sigma} p_{i,\sigma}^2, $$
and thus
\begin{equation}\label{eq:overview-basic}
    \frac{1}{n}\sum_{i,\sigma} p_{i,\sigma}^2 \;\le\; \alpha + \frac{1}{|L|}.
\end{equation}

We now use the fact that the elements of $L$ all have high agreement with $r$ to lower bound the left-hand side of the above inequality.
Without a hint, the sum $\frac{1}{n}\sum_i p_{i,r_i}$ is at least $\beta$, and by the Cauchy-Schwarz inequality, $\frac{1}{n}\sum_i p_{i,r_i}^2 \ge \beta^2$.
Combined with~\eqref{eq:overview-basic}, this gives $\beta^2 \le \alpha + 1/|L|$, recovering the Johnson bound $\beta^* = \sqrt{\alpha}$.

Crucially, the Cauchy-Schwarz lower bound is tight when all of the $p_{i, r_i}$ are equal to $\beta$, i.e., when the list's agreement with $r$ is spread uniformly across coordinates.
The introduction of a hint can prevent this uniform distribution from being feasible, leading to a stronger bound.
Suppose we have a hint $h$ with $\agree(r,h) = \eta$.
We partition the coordinates into $U$ (where $r$ and $h$ agree, $|U| = \eta n$) and $V = [n] \setminus U$.
Define macroscopic variables $a = \frac{1}{n}\sum_{i \in U} p_{i,r_i}$ (the average agreement coming from $U$) and $b = \frac{1}{n}\sum_{i \in V} p_{i,r_i}$ (the average agreement coming from $V$), so that $a + b \geq \beta$.
Since on $U$ we have $r_i = h_i$, the quantity $a$ also lower bounds the list's average agreement with $h$.
Since $h$ is an element of the code, this gives the capacity constraint $a \le \alpha$.
Applying Cauchy-Schwarz separately to $U$ and $V$ yields
\begin{equation}\label{overview:single-hint}
\frac{a^2}{\eta} + \frac{b^2}{1-\eta} \le \alpha + 1/|L|.
\end{equation}
When $\eta > \sqrt{\alpha}$, the capacity constraint $a \le \alpha$ prevents the variables from achieving the unconstrained optimum, forcing an uneven split between $U$ and $V$ that increases the quadratic sum.
This gives a list-decoding threshold $\beta^* < \sqrt{\alpha}$, defined as the maximum of $a + b$ subject to the quadratic and capacity constraints (Optimization~\ref{opt:johnson-large-alphabet-single-hint}).

To get the quantitative bound $|L| = O(1/\epsilon)$ when $\beta = \beta^* + \epsilon$, we need to show the LHS of~\eqref{overview:single-hint} is $\alpha + \Omega(\epsilon)$.
This can be done by hand (e.g., by calculating gradients at the optimum) for the current single-hint, large-alphabet case.
In the more general cases, however, this can be done using Lagrangian duality, which gives tools for understanding how the optimum of a convex optimization
changes under perturbations of the constraints.
Specifically, we use the relaxed Slater condition to show that strong duality holds, allowing us to work with the Lagrangian and perform a first-order analysis of it.
We do this in Lemma~\ref{lem:slater-point} (which shows that the optimization satisfies the relaxed Slater condition) and Lemma~\ref{lem:generalized-first-order} (which does the first-order analysis).

The other cases follow the same template with several modifications.

\paragraph{Addressing smaller alphabets.}
Let's again start by recovering the standard small-alphabet Johnson bound (without a hint).
Recall that in the large-alphabet case, we lower bounded the sum in \eqref{eq:overview-basic} by ignoring all the $p_{i, \sigma}$ for $\sigma \neq r_i$ and applying Cauchy-Schwarz.
When the alphabet size $q$ is small, the terms that we ignored are no longer negligible.
Write $\beta$ for the average agreement $\frac{1}{n}\sum_i p_{i,r_i}$.
Applying Cauchy-Schwarz gives $\frac{1}{n}\sum_i p_{i,r_i}^2 \ge \beta^2$.
For the non-received symbols, the total mass is $\frac{1}{n}\sum_{i, \sigma \neq r_i} p_{i,\sigma} = 1 - \beta$, so Cauchy-Schwarz gives
\[
    \frac{1}{n}\sum_i \sum_{\sigma \neq r_i} p_{i,\sigma}^2 \ge \frac{(1-\beta)^2}{q-1}.
\]
Thus,
\[
    \frac{1}{n}\sum_{i,\sigma} p_{i,\sigma}^2 \;\ge\; \beta^2 + \frac{(1-\beta)^2}{q-1}.
\]
Substituting this lower bound into~\eqref{eq:overview-basic} and solving for $\beta$ recovers the standard small-alphabet Johnson bound.

When a hint is present, we partition the sum $\sum_{i,\sigma} p_{i,\sigma}^2$ more finely.
Within $U$ (coordinates where $r_i = h_i$), there are two groups: $\sigma = r_i$ and $\sigma \neq r_i$.
Within $V$ (coordinates where $r_i \neq h_i$), there are three groups: $\sigma = r_i$, $\sigma = h_i$, and $\sigma$ equal to neither.
As in the large-alphabet case, we introduce macroscopic variables corresponding to this partition and derive constraints based on the agreement between list elements (the ellipsoid constraint) and the agreement between the hint and list elements (the capacity constraint).
This yields Optimization~\ref{opt:johnson-small-alphabet-single-hint}.

\paragraph{Addressing multiple hints.}
When there are multiple hints, $h_1,\ldots,h_m$, the hints can differ not only in their agreement with the received word but also in their agreement with one another.
To account for this, we take a finer partition of coordinates.
Each coordinate $i$ is classified by the full pattern of symbol equalities among the received word and all hints, forming a partition $\pi_i$ of $\{0, 1, \ldots, m\}$ (where $0$ represents $r$).
Coordinates with the same partition type are grouped together, and we define macroscopic variables $a_{\pi, S}$ tracking the normalized sum of $p_{i,\sigma}$ for each part $S$ of each partition $\pi$, along with a residual variable $b_\pi$ for the remaining symbols.
Applying Cauchy-Schwarz within each group gives an ellipsoid bound similar to \eqref{overview:single-hint} where the LHS is a weighted sum of squares of the variables and the RHS is $\alpha + 1/|L|$ (the ellipsoid constraint), while the hint agreements impose linear capacity constraints.
This yields Optimization~\ref{opt:multi}, whose optimum $\beta^*$ is the list-decoding threshold.

\subsubsection{Lower bounds on list size}

To show tightness, we prove that for any $\beta = \beta^* - \epsilon$ there exist codes with exponentially many codewords all having agreement at least $\beta$ with $r$.

First, recall the standard proof of the tightness of the Johnson agreement $\sqrt{\alpha}$ (without hints).
We take the received word $r$ to be the all-zeros word. Now we create a code with maximum agreement $\alpha$ and with exponentially many codewords having agreement $\sqrt{\alpha}-\epsilon$ with $r$. This is done by the probabilistic method: the codewords are chosen independently, with a random $(\sqrt{\alpha}-\epsilon)$-fraction of coordinates set to $0$ and the remaining coordinates chosen independently and uniformly from $\Sigma$. Their agreement with $r$ is guaranteed by construction, and their pairwise maximum agreement follows from exponentially good concentration bounds (and a union bound over all pairs of codewords).

We now describe the construction for the single-hint, large-alphabet case; the general case (Section~\ref{sec:general}) follows the same template with a finer partition of coordinates.

As above, we will take $r$ to be the all-zeros word.
We have to create a code $C$ and a hint $h \in C$.
The hint $h$ is chosen to be $0$ (and hence agree with $r$) on the first $\eta n$ coordinates and to be $1$ on the remaining coordinates.
Partition $[n]$ into blocks $U$ (where $r$ and $h$ agree) and $V$ (where they disagree).

We now create the rest of the code $C$ with exponentially many codewords close to $r$. We have to balance this with the fact that (1) all the new codewords should have agreement at most $\alpha$ with each other and (2) all the new codewords have to have agreement at most $\alpha$ with $h$. To do this, we again use the probabilistic method, this time guided by
an optimal solution $(a^*, b^*)$ to Optimization~\ref{opt:johnson-large-alphabet-single-hint}.

After a mild scaling, consider the feasible solution $(a, b) = (1-\epsilon/\beta^*)(a^*, b^*)$ with objective $a + b = \beta = \beta^* - \epsilon$ and strict slack $\delta = \Theta(\epsilon)$ in the ellipsoid constraint.

We now take $C$ to contain exponentially many independently picked codewords $c$, uniformly at random subject to
having $an$ 0s among the $U$ coordinates and $bn$ 0s among the $V$ coordinates. (Thus, the remaining coordinates are picked uniformly at random from $\Sigma \setminus \{0,1\}$.)
The pairwise agreement of two codewords picked from this distribution on the $U$ coordinates and the $V$ coordinates will be $(a/\eta)^2 + \delta/2$ and $(b/(1-\eta))^2 + \delta/2$, respectively, with high probability,
and thus the total agreement will be
$$a^2/\eta + b^2/(1-\eta) + \delta/2 \le \alpha - \delta/2 < \alpha,$$
where the first inequality follows from the fact that $(a,b)$ is feasible for
Optimization~\ref{opt:johnson-large-alphabet-single-hint}.
(Note that in the large-alphabet setting, the collisions on symbols not equal to $0$ are negligible and can be rolled into the $\delta$ term.)

Furthermore, notice that the agreement between $h$ and a codeword picked from this distribution is exactly $a$, which is at most $\alpha$ since $a$ is feasible.
Finally, every codeword in $C \setminus \{h\}$ has exactly $an + bn = \beta n$ 0s and hence agreement exactly $\beta$ with $r$.

\subsubsection{Decoding algorithm for RS codes}

Since Reed-Solomon codes have $q \ge n$, they fall into the large-alphabet regime.
We are given a received word $r$ and $m$ hints $h_1, \ldots, h_m \in C$.
In this regime, we partition the coordinates based on which subset $S \subseteq [m]$ of hints agree with $r$ at that coordinate: $U_S$ is the set of coordinates $i$ where the set of hints that agree with $r$ on coordinate $i$ is $S$.
Let $\eta_S = |U_S|/n$, which captures the size of the part $U_S$.

At a high level, the algorithm for efficiently list-decoding Reed-Solomon codes from
agreement $\beta$ will use the weighted Guruswami-Sudan list-decoding algorithm~\cite{GuruswamiSudan,GuruswamiThesis} with weights derived
from an optimal solution $(a_S^*)_{S \subseteq [m]}$ to Optimization~\ref{opt:multi-hint-large-alphabet}.

The weighted GS list-decoder takes a weight vector $w \in \mathbb{R}_{\ge 0}^n$ and returns all codewords $c$ such that the weighted agreement between $c$ and $r$ exceeds the threshold $\sqrt{\alpha n \sum_{i=1}^n w_i^2} + \epsilon w_{\max}$.

We set the weight for all the coordinates $i\in U_S$ to be the same:
$$w_i = a_S^*/\eta_S.$$
With this choice of weights (and using the feasibility of the $a_S^*$), the Guruswami-Sudan decoding threshold simplifies to $\alpha n$.
For any codeword $c$ with $\agree(c,r) \ge \beta^* + \epsilon$, let $a_S = \frac{1}{n}|\{i \in U_S: r_i = c_i\}| $.
Then the weighted agreement between $c$ and $r$ equals
$$n \sum_{S} \frac{a_S^* a_S}{\eta_S},$$
which, based on the properties of Optimization~\ref{opt:multi-hint-large-alphabet}, can be shown to exceed $\alpha n$.
Thus the codeword $c$ will be found by our invocation of the weighted Guruswami-Sudan decoder.






\subsubsection{Organization}
We begin in Section~\ref{sec:large-q-single} with the simplest case---a single hint over a large alphabet---which contains the main ideas.
The next three sections extend the analysis to small alphabets (Section~\ref{sec:small-q}), multiple hints over a large alphabet (Section~\ref{sec:multi-large-q}), and the fully general case (Section~\ref{sec:general}). Since the first two are special cases of the general setting, we focus on the additional complications each introduces and the resulting optimization problem, deferring the full proof to Section~\ref{sec:general}. Section~\ref{sec:decoding} shows how to make these bounds algorithmic for Reed-Solomon codes using the weighted Guruswami-Sudan algorithm.

\subsection*{AI use}
The results were obtained without any AI involvement. The writing was done without
any AI involvement. We used AI for computing the optimum of our optimization for assorted instantiations, and for feedback on our writeup.

\subsection*{Acknowledgements}
We are grateful to the anonymous reviewers of FOCS 2026 and SODA 2027 for valuable comments and corrections. SK thanks John Kim for valuable discussions.

\section{Preliminaries and Notation}

Let $\Sigma$ be an alphabet of size $q = |\Sigma|$, and let $\Sigma^n$ be the set of words of length $n$ over $\Sigma$. Throughout, $n$ denotes the codeword length and $m$ the number of hints; we think of $n$ as growing and $m$ as fixed unless otherwise stated.

For $x, y \in \Sigma^n$, define the \emph{Hamming agreement} $\Agree(x, y) = |\{i \in [n] : x_i = y_i\}|$ and the \emph{normalized agreement} $\agree(x, y) = \Agree(x, y)/n$.
A code $C \subset \Sigma^n$ has \emph{maximum agreement} $\alpha$ if $\agree(x, y) \leq \alpha$ for all distinct $x, y \in C$.
For a subset $S \subseteq [n]$, we write $\Agree_S(x, y) = |\{i \in S : x_i = y_i\}|$ for the agreement restricted to coordinates in $S$.

We use $\circ$ to denote concatenation of strings.

\subsection{Random Codes with a Certain Type}

In our lower bounds, we construct codes in which each alphabet symbol occurs an exact number of times in every codeword.
A \emph{type} specifies this distribution of symbols.

\begin{definition}[Type]
    Let $t : \Sigma \to \N$ be such that $\sum_{\sigma \in \Sigma} t(\sigma) = n$.
    A string $x \in \Sigma^n$ has type $t$ if for each $\sigma \in \Sigma$, the number of occurrences of $\sigma$ in $x$ is exactly $t(\sigma)$.
\end{definition}

Exponentially large codes of a prescribed type exist; this can be proved using the method of averaged bounded differences (see, e.g.,~\cite{dubhashiConcentrationMeasureAnalysis2009}).

\begin{lemma} \label{lem:random-type}
    Fix an alphabet $\Sigma$, $n \in \mathbb{N}$, a type $t: \Sigma \to \mathbb{N}$, and a constant $\epsilon > 0$ independent of $n$.
    If
    $$
        \alpha \geq \sum_{\sigma \in \Sigma} \left(\frac{t(\sigma)}{n}\right)^2 + \epsilon,
        \qquad \text{and} \qquad
        \ell^2\exp(-\epsilon^2 n/2) < 1,
    $$
	    then there exists a code of size $\ell$ in which every codeword has type $t$ and the maximum agreement is at most $\alpha$.
    In particular, we can take $\ell = \exp(\Omega(n))$.
\end{lemma}

\subsection{Expected agreement}

We will need the following basic bound on the expected agreement between two distinct elements of a set.

\begin{lemma}\label{lem:expected-agreement-distinct}
    Let $L \subset \Sigma^n$.
    Let $p_{i, \sigma} = \Pr_{x \sim L}[x_i = \sigma]$.
    Then
    $$
        \E_{x, y \sim L, x \neq y}[\agree(x, y)] \geq \left(\frac{1}{n} \sum_{i=1}^n \sum_{\sigma \in \Sigma} p_{i,\sigma}^2\right) - \frac{1}{|L|}.
    $$
\end{lemma}

This follows immediately from the next two claims.

\begin{claim}
    Let $L \subset \Sigma^n$.
    Let $p_{i, \sigma} = \Pr_{x \sim L}[x_i = \sigma]$.
    Then
    $$
        \E_{x, y \sim L}[\agree(x, y)] = \frac{1}{n} \sum_{i=1}^n \sum_{\sigma \in \Sigma} p_{i,\sigma}^2.
    $$
\end{claim}

\begin{proof}
    \begin{align*}
        \E_{x, y \sim L}[\agree(x, y)] & = \E_{x, y \sim L, i \sim [n]}\1_{x_i = y_i}                                                                                 \\
                                       & = \E_{x, y \sim L, i \sim [n]}\left[\sum_{\sigma \in \Sigma}\1_{x_i = y_i = \sigma}\right]                                             \\
                                       & = \E_{i \sim [n]} \sum_{\sigma \in \Sigma} \left[ \E_{x, y \sim L}\left[\1_{x_i = y_i = \sigma}\right] \right] & (\text{Linearity})    \\
                                       & = \E_{i \sim [n]} \sum_{\sigma \in \Sigma} \left[ \Pr_{x, y \sim L}\left[x_i = y_i = \sigma\right] \right]                             \\
                                       & = \E_{i \sim [n]} \sum_{\sigma \in \Sigma} \left[ \Pr_{x \sim L}\left[x_i = \sigma\right]^2 \right]            & (\text{Independence}) \\
                                       & = \E_{i \sim [n]} \sum_{\sigma \in \Sigma}  p_{i,\sigma}^2.
    \end{align*}
\end{proof}

\begin{claim}
    Let $L \subset \Sigma^n$.
    Then
    \[
        \E_{c \neq c' \sim L}[\agree(c, c')] \geq \E_{c, c' \sim L}[\agree(c, c')] - \frac{1}{|L|}.
    \]
\end{claim}

\begin{proof}
    Let $\ell \coloneqq |L|$.
    Sample $(c,c')$ by drawing two elements from $L$ independently and uniformly at random with replacement.
    Then
    \[
        \E_{c,c' \sim L}[\agree(c,c')]
        = \Pr[c=c']\,\E[\agree(c,c')\mid c=c']
        + \Pr[c\ne c']\,\E[\agree(c,c')\mid c\ne c'].
    \]
    When $c=c'$, we have $\agree(c,c')=1$.
    Also $\Pr[c=c']=1/\ell$.
    Denote
    \(
    A_{\ne} \coloneqq \E[\agree(c,c')\mid c\ne c'] = \E_{c\ne c'\sim L}[\agree(c,c')].
    \)
    Therefore
    \[
        \E_{c,c' \sim L}[\agree(c,c')] = \frac{1}{\ell}\cdot 1 + \Big(1-\frac{1}{\ell}\Big) A_{\ne}.
    \]
    Rearranging gives
    \[
        A_{\ne} = \frac{\E_{c,c' \sim L}[\agree(c,c')] - 1/\ell}{1 - 1/\ell} \ge \E_{c,c' \sim L}[\agree(c,c')] - \frac{1}{\ell},
    \]
    where the inequality uses that $1/(1-1/\ell) \ge 1$.
    This proves the claim.
\end{proof}

\subsection{Convex Optimization and Duality}

We show that the list-decoding threshold is the optimum of a convex program.
Our proofs of the upper bound on list size rely heavily on several properties of such programs.
See \cite{Boyd} for a reference.

Our convex programs take the following form:
\begin{align*}
    \beta^* \;=\; \max_{x \in \mathcal{X}} \quad & c^T x \\
    \text{s.t.} \quad & f(x) \le \alpha,
\end{align*}
where $f$ is a convex and differentiable function, $\mathcal{X}$ is a non-empty convex set defined by linear constraints, and $c$ is a fixed vector.
The \emph{Lagrangian} of this program, with respect to the convex constraint $f(x) \le \alpha$, is
\[
    \mathcal{L}(x, \lambda) \;=\; c^T x - \lambda\bigl(f(x) - \alpha\bigr),
\]
where $\lambda \ge 0$ is a dual variable (or \emph{KKT multiplier}) associated with the constraint $f(x) \le \alpha$.
Weak duality asserts that
$
    \beta^* \;\le\; \min_{\lambda \geq 0}\max_{x \in \mathcal{X}} \;\mathcal{L}(x, \lambda).
$
\emph{Strong duality} holds when this inequality is tight, i.e., when there exists $\lambda^* \ge 0$ such that
\[
    \beta^* \;=\; \max_{x \in \mathcal{X}} \;\mathcal{L}(x, \lambda^*).
\]
When $\lambda^* > 0$, \emph{complementary slackness} further requires that $f(x^*) = \alpha$ at any primal optimum $x^*$.

The relaxed Slater condition provides a convenient sufficient condition for strong duality: if there exists a point $z \in \mathcal{X}$ satisfying $f(z) < \alpha$ (i.e., a \emph{strictly} feasible point for the convex constraint), then strong duality holds.
We call this point the \emph{Slater point}.
In each of our optimization problems, we identify such a Slater point explicitly.

\subsection{List decoding with a hint}

We consider four settings, depending on whether one or more hints are given and whether the alphabet is large or small.
In the large-alphabet setting, $q = |\Sigma|$ is assumed to be sufficiently large as a function of the other parameters. The small-alphabet setting treats $q$ as an explicit parameter.
These two regimes are analogous to the large-alphabet\footnote{The large-alphabet Johnson bound is sometimes called the alphabet-free Johnson bound.} and small-alphabet versions of the classical Johnson bound.

The problem is governed by the following parameters:
	\begin{itemize}
	    \item $\alpha \in [0,1]$, the maximum agreement of the code $C$.
	    \item $\eta$, capturing the relationship between the hints and the received word. For a single hint, $\eta = \agree(h, r)$. For multiple hints, $\eta$ must also capture the relationship between the hints. The definition there is slightly more involved and is given in the relevant sections.
	    \item $m$, the number of hints.
	    \item $q = |\Sigma|$, the alphabet size.
	\end{itemize}

An instance of the list-decoding problem with hints is characterized by a code $C \subseteq \Sigma^n$, a received word $r \in \Sigma^n$, a collection of one or more hints, and an alphabet size $q$.
We write $P$ for the tuple of parameters characterizing a given setting (e.g., $P = (\alpha, \eta)$ for a single hint over a large alphabet).

We call an agreement profile $\eta$ $(q,\alpha)$-realizable if there exist a blocklength $n$, an alphabet $\Sigma$ of size $q$, a received word $r\in\Sigma^n$, and hints $h_1,\ldots,h_m\in\Sigma^n$ having agreement profile $\eta$ such that $\agree(h_j,h_k)\leq\alpha$ for every pair of distinct hints $j,k\in[m]$. Equivalently, the hints are contained in a code of maximum agreement at most $\alpha$. In the large-alphabet setting, we simply say that $\eta$ is $\alpha$-realizable if it is $(q,\alpha)$-realizable for some $q$; it is then realizable over every larger alphabet by embedding.

\begin{definition}[(Average) list-decodability]\label{def:average-list-decodability}
We say $C \subseteq \Sigma^n$ is $(\beta, \ell)$-list-decodable at $r$ if every collection $L \subseteq C$ of $\ell$ codewords
contains some $c \in L$ such that
$$ \agree(c,r) < \beta.$$

We say $C \subseteq \Sigma^n$ is $(\beta, \ell)$-average-list-decodable at $r$ if for every collection $L \subseteq C$ of $\ell$ codewords,
we have
$${\mathbb E}_{c \in L} [\agree(c, r) ] < \beta.$$
\end{definition}

\begin{definition}[Average list-decoding threshold]\label{def:list-decoding-radius}
    The \emph{average list-decoding threshold} for a list-decoding problem with parameters $P$ is a value $\beta^*(P) \in [0,1]$ such that for every $\epsilon > 0$:
    \begin{itemize}
	        \item There is a constant $C_P$ depending only on $P$ such that
        if $\beta = \beta^*(P) + \epsilon$, then in every instance with parameters $P$,
        $C$ is $(\beta,\frac{C_P}{\epsilon})$-average-list-decodable at $r$.
        \item If $\beta \leq \beta^*(P) - \epsilon$ for any $\epsilon > 0$, then there exists an instance with parameters $P$ such that $C$ is {\bf not} $(\beta, \exp(\Omega(n)))$-list-decodable.
    \end{itemize}
	    In the large-alphabet case, the lower bound (the second bullet point) is only required to hold when the alphabet size is sufficiently large as a function of the other parameters and $1/\epsilon$. More specifically, our lower bounds hold so long as the alphabet has size at least $m+1 + \ceil{\frac{4}{\epsilon \alpha}}$.
\end{definition}

\begin{remark}
    Average list decodability implies standard list decodability.
    Also note that the negative direction in the above definition asks for large list sizes for standard list-decodability, not average list decodability.
\end{remark}

\section{Johnson Bound with a hint, \texorpdfstring{$q$ large}{q large}} \label{sec:large-q-single}

The list-decoding threshold in this setting is captured by the following optimization problem.
\footnote{
When a denominator such as $\eta$ or $1-\eta$ is zero, we use the convention that the corresponding variable is forced to be zero and the associated quadratic term is omitted.
We use the same convention for $\eta_S$ and $\eta_\pi$, which appear in later sections.
}

\begin{optimization}\label{opt:johnson-large-alphabet-single-hint}
    $\alpha, \eta \in [0,1]$ are fixed parameters, and the decision variables are $a, b$.
    \begin{align*}
        \max \quad        & a + b \\
        \text{s.t.} \quad & \frac{a^2}{\eta} + \frac{b^2}{1 - \eta} \leq \alpha                        \\
                          & 0 \leq a \leq \min(\eta, \alpha)                             \\
                          & 0 \leq b \leq 1 - \eta
    \end{align*}
\end{optimization}

\begin{theorem}\label{thm:large-alphabet-johnson-single}
    Consider the list-decoding problem with a single hint. Let $\eta \in [0,1]$ be the agreement between the hint and the received word, and let $\alpha \in (0,1)$ be the maximum agreement of the code.
    Let $\beta^*(\alpha, \eta)$ be the optimum of Optimization~\ref{opt:johnson-large-alphabet-single-hint}.
	    Then $\beta^*(\alpha, \eta)$ is the average list-decoding threshold for the one-hint, large-alphabet list-decoding problem.
\end{theorem}

	\begin{proof}[Proof of the upper bound ($\beta > \beta^*$)]
	    Let $C$ be a code of maximum agreement $\alpha$ over a large alphabet $\Sigma$.
	    Let $r \in \Sigma^n$ be a received word and $h \in C$ a hint such that $\agree(r, h) = \eta$.
	    Let $\epsilon > 0$, and let $\beta = \beta^*(\alpha, \eta) + \epsilon$ be the target agreement.
	    Suppose $L \subset C \setminus \{h\}$ is a list of codewords such that $\E_{c \sim L}[\agree(c, r)] \geq \beta$.
	    We show that $|L| \leq \frac{1}{\epsilon \alpha}$.

    Let $p_{i,\sigma} \coloneqq \Pr_{c\sim L}[c_i=\sigma]$ be the fraction of codewords in the list that place symbol $\sigma$ at coordinate $i$.
    By Lemma~\ref{lem:expected-agreement-distinct}, we have the following inequality.
    \begin{equation}\label{eq:expected_agreement_bound}
        \left(\frac{1}{n}\sum_{i=1}^n \sum_{\sigma\in\Sigma} p_{i,\sigma}^2\right) - \frac{1}{|L|} \leq \alpha.
    \end{equation}

    Let $U \subset [n]$ be the set of coordinates where $r$ and $h$ agree, and note that $|U| = \eta n$.
    Define
    \[
        a \coloneqq \frac{1}{n}\sum_{i\in U} p_{i,r_i}, \qquad b \coloneqq \frac{1}{n}\sum_{i\notin U} p_{i,r_i}.
    \]

    Since the average agreement of codewords in $L$ with $r$ is at least $\beta$, linearity of expectation gives $a + b = \E_{c \sim L}[\agree(c, r)] \ge \beta$.

    We claim that $a, b$ satisfy all of the linear constraints of Optimization~\ref{opt:johnson-large-alphabet-single-hint}.
    For the $a \leq \alpha$ constraint, note that
    $$a = \frac{1}{n}\sum_{i\in U} p_{i,r_i} = \frac{1}{n}\sum_{i\in U} p_{i,h_i} \leq \frac{1}{n}\sum_{i=1}^n p_{i,h_i}.$$
	    Since the RHS of the above is the average agreement of codewords in $L$ with $h$, and since every codeword in $C$ has agreement at most $\alpha$ with $h$, we have $a \leq \alpha$. The constraints $a \le \eta$ and $b \le 1-\eta$ follow trivially from the definitions.

	    Next, we lower bound the sum in~\eqref{eq:expected_agreement_bound}.
	    First, drop all the terms $p_{i, \sigma}$ where $\sigma \neq r_i$.
	    Then split the remaining sum over $U$ and its complement and apply Cauchy-Schwarz to each part. Thus, we have
	    \begin{equation} \label{eq:e1}
	        \frac{a^2}{\eta} + \frac{b^2}{1-\eta}  \;\le\; \alpha + \frac{1}{|L|}.
	    \end{equation}

    Define $f(a,b) \coloneqq \frac{a^2}{\eta} + \frac{b^2}{1-\eta}$.
    We show that $f(a,b) \ge \alpha + \Omega(\epsilon)$, which, combined with~\eqref{eq:e1}, gives $|L| = O(1/\epsilon)$.

	    Let $\mathcal{X}$ denote the polyhedron defined by the linear constraints of the optimization (i.e., all the constraints except for $f(a, b) \leq \alpha$).
	    Let $x^* = (a^*, b^*)$ be the optimal solution to Optimization~\ref{opt:johnson-large-alphabet-single-hint}.
	    Since $z = (0,0) \in \mathcal{X}$ satisfies $f(z) = 0 < \alpha$, strong duality holds by Slater's relaxed condition.
	    Thus, there exists a KKT multiplier $\lambda \ge 0$ such that $x^*$ maximizes the Lagrangian
	    \[
	        \mathcal{L}(a, b) \;=\; (a + b) - \lambda\bigl(f(a,b) - \alpha\bigr)
	    \]
    over $\mathcal{X}$, with complementary slackness $\lambda(f(x^*) - \alpha) = 0$.
    If $\lambda = 0$, then $x^*$ maximizes $a + b$ over $\mathcal{X}$ unconditionally, contradicting $a + b > \beta^* = a^* + b^*$ for our $x = (a,b) \in \mathcal{X}$. So $\lambda > 0$. Additionally, we have $f(x^*) = \alpha$ by complementary slackness.
    By the first-order condition for convexity of $f$,
    \[
        f(x) \;\ge\; f(x^*) + \langle \nabla f(x^*),\, x - x^* \rangle
        \;=\; \alpha + \langle \nabla f(x^*),\, x - x^* \rangle.
    \]
    So it suffices to show $\langle \nabla f(x^*),\, x - x^* \rangle \ge \epsilon \alpha$.
    Since $x^*$ maximizes $\mathcal{L}$ over $\mathcal{X}$ and $x \in \mathcal{X}$, the first-order optimality condition gives $\langle \nabla \mathcal{L}(x^*),\, x - x^* \rangle \le 0$.
    Expanding $\nabla \mathcal{L}(x^*) = (1,1) - \lambda \nabla f(x^*)$ yields
    \[
        \epsilon \;\le\; (a+b) - (a^*+b^*) \;\le\; \lambda\,\langle \nabla f(x^*),\, x - x^* \rangle.
    \]
    To bound $\lambda$, use $\mathcal{L}(z) \le \mathcal{L}(x^*)$, which gives $\lambda\alpha \le \beta^*$, so $\lambda \le \beta^*/\alpha$.
    Combining, we get $\langle \nabla f(x^*),\, x - x^* \rangle \ge \epsilon/\lambda \ge \epsilon\alpha/\beta^* \geq \epsilon\alpha$ as required.
\end{proof}

\begin{remark}[List size]
    The proof gives the quantitative bound
    \[
        |L| \le \frac{\beta^*}{\alpha(\beta-\beta^*)}.
    \]
\end{remark}

\begin{proof}[Proof of the lower bound ($\beta < \beta^*$)]
    Fix $\epsilon > 0$ and suppose $\beta = \beta^*(\alpha, \eta) - \epsilon$.
    Let $(a^*, b^*)$ be an optimal solution to Optimization~\ref{opt:johnson-large-alphabet-single-hint}, so that $a^* + b^* = \beta^*$ and $(a^*, b^*)$ satisfies all constraints.
    Define $(a, b) \coloneqq (1 - \epsilon/\beta^*)\,(a^*, b^*)$.
    Then $a + b = \beta^* - \epsilon = \beta$, and the linear constraints remain satisfied since we are scaling down.
    For the ellipsoid constraint, by the scaling of $(a,b)$ we have $f(a,b) = (1 - \epsilon/\beta^*)^2 f(a^*, b^*)$.
    Since $\beta^* \le 1$, the factor $(1 - \epsilon/\beta^*)^2 \le (1-\epsilon)^2 \le 1 - \epsilon$.
    Furthermore, $f(a^*, b^*) \leq \alpha$.
    Combining these inequalities, we obtain $f(a,b) \le (1 - \epsilon)\alpha = \alpha - \delta$, where $\delta \coloneqq \epsilon\alpha$.
    Assume $|\Sigma| \ge 2 + \ceil{4/\delta}$, and write $\Sigma = \{0,1\} \cup \Gamma$ where $|\Gamma| \ge \ceil{4/\delta}$.

    We construct a received word $r$, a hint $h$, and a code $L$ of size $\exp(\Omega(n))$ such that every codeword in $L$ has agreement at least $\beta$ with $r$, and $L \cup \{h\}$ has maximum pairwise agreement at most $\alpha$.

    Let $r$ be the all-zeros word.
    Let $h$ be the word with the first $\eta n$ coordinates equal to $0$ and the remaining $(1-\eta)n$ coordinates equal to $1$.
    Let $U = \{1,\ldots,\eta n\}$ and $V = [n] \setminus U$, so that $U$ is the set of coordinates where $r$ and $h$ agree.
    We build $L$ as follows so that each codeword has exactly $an$ zeros on $U$ and exactly $bn$ zeros on $V$.\footnote{Fixing the type ensures that the agreement of every codeword in $L$ with the hint $h$ is exactly $a \le \alpha$; sampling coordinates independently could produce codewords exceeding the agreement bound with non-negligible probability.}

    Define a type $t_U$ over $\Sigma$ of length $|U|$ by $t_U(0) = an$, $t_U(1) = 0$, and $t_U(\gamma) = (|U| - an)/|\Gamma|$ for each $\gamma \in \Gamma$.
    By Lemma~\ref{lem:random-type}, there exists a code $L_U\subset \Sigma^{U}$ of size $\exp(\Omega(|U|))$, all of whose codewords have type $t_U$, and whose maximum agreement (normalized by $|U|$) is at most
    \[
        \sum_{\sigma \in \Sigma} \Big(\frac{t_U(\sigma)}{|U|}\Big)^2 + \frac{\delta}{4}
        \;=\; \Big(\frac{a}{\eta}\Big)^2 + \frac{(1- a/\eta)^2}{|\Gamma|} + \frac{\delta}{4}
        \;\le\; \Big(\frac{a}{\eta}\Big)^2 + \frac{\delta}{2},
    \]
    where the last inequality uses $|\Gamma| \ge 4/\delta$.

    Similarly, define a type $t_V$ over $\Sigma$ of length $|V|$ by setting $t_V(0)=bn$, $t_V(1) = 0$, and spreading the remaining $|V|-bn$ symbols evenly across $\Gamma$.
    By Lemma~\ref{lem:random-type}, there exists $L_V\subset \Sigma^{V}$ of size $\exp(\Omega(|V|))$ with maximum agreement (normalized by $|V|$) at most $(b/(1-\eta))^2 + \delta/2$.

    Suppose one block, say $V$, has size $o(n)$. Then $1-\eta = o(1)$, hence any agreement on $V$ is $o(1)$ and we can define a code with the desired properties by taking $L_U$ and setting the coordinates in $V$ to an arbitrary value $\sigma \in \Gamma$. Since the agreement in $V$ is $o(1)$, this can be absorbed in the constant slack $\delta$.

    Thus, assume both blocks have size $\Omega(n)$.
    Let $N \coloneqq \min(|L_U|, |L_V|) = \exp(\Omega(n))$, and fix arbitrary enumerations $L_U = \{u_1, \ldots\}$ and $L_V = \{v_1, \ldots\}$.
    Set $L \coloneqq \{u_j \circ v_j : j \in [N]\}$, so $|L| = \exp(\Omega(n))$.

    It remains to verify the required properties.
    First, we show that $L \cup \{h\}$ has maximum agreement at most $\alpha$.
    For distinct $c=u_j\circ v_j, \; c'=u_k\circ v_k$ in $L$ (with $j \neq k$, so $u_j \neq u_k$ and $v_j \neq v_k$),
    \begin{align*}
        \agree(c,c')
         & = \frac{1}{n}\bigl( \Agree_{U}(u,u') + \Agree_{V}(v,v') \bigr)                                                                                     \\
         & \le \eta\Big(\Big(\frac{a}{\eta}\Big)^2 + \frac{\delta}{2}\Big) + (1-\eta)\Big(\Big(\frac{b}{1-\eta}\Big)^2 + \frac{\delta}{2}\Big) \\
         & = \frac{a^2}{\eta} + \frac{b^2}{1-\eta} + \frac{\delta}{2}
        \;\le\; \alpha - \frac{\delta}{2}.
    \end{align*}
    For any $c \in L$, its agreement with the hint is $\agree(c,h) = a \le \alpha$, since each codeword in $L$ has exactly $an$ zeros on $U$ (where $h$ is also zero) and no ones on $V$ (where $h$ equals $1$).
    Finally, each $c \in L$ satisfies $\agree(c,r) = a + b = \beta$.
\end{proof}

Next, we solve Optimization~\ref{opt:johnson-large-alphabet-single-hint} analytically to find the exact threshold $\beta^*$.

\begin{theorem}\label{thm:solve-single-opt}
    Fix parameters $(\alpha,\eta) \in [0,1]^2$.
    Define the critical threshold
    \[
        \eta_{\mathrm{crit}} \coloneqq \frac{(1-\alpha) + \sqrt{1 - 2\alpha + 5\alpha^2}}{2}.
    \]
	    The exact average list-decoding threshold $\beta^*=\beta^*(\alpha,\eta)$ of Optimization~\ref{opt:johnson-large-alphabet-single-hint} is
    \[
        \beta^*
        = \begin{cases}
            \sqrt{\alpha},                                                   & \eta \leq \sqrt{\alpha}, \\
            \alpha + \sqrt{(1-\eta)\Bigl(\alpha-\frac{\alpha^2}{\eta}\Bigr)}, & \sqrt{\alpha} < \eta \leq \eta_{\mathrm{crit}}, \\
            \alpha + 1 - \eta,                                               & \eta > \eta_{\mathrm{crit}}.
        \end{cases}
    \]
    Moreover, the unique optimal solution is
    \[
        (a^*,b^*) = \begin{cases}
            (\eta\sqrt{\alpha}, (1-\eta)\sqrt{\alpha}), & \eta \leq \sqrt{\alpha}, \\
            \Bigl(\alpha, \sqrt{(1-\eta)\Bigl(\alpha-\frac{\alpha^2}{\eta}\Bigr)}\Bigr),     & \sqrt{\alpha} < \eta \leq \eta_{\mathrm{crit}}, \\
            (\alpha, 1 - \eta),                           & \eta > \eta_{\mathrm{crit}}.
        \end{cases}
    \]
\end{theorem}

\begin{proof}[Proof of Theorem~\ref{thm:solve-single-opt}]
    We first solve the relaxation that ignores all linear constraints, keeping only the ellipsoid $\frac{a^2}{\eta}+\frac{b^2}{1-\eta} \le \alpha$ and $a,b \ge 0$.
    By Cauchy-Schwarz, $(a+b)^2 \le \bigl(\frac{a^2}{\eta}+\frac{b^2}{1-\eta}\bigr)(\eta + (1-\eta)) \le \alpha$, with equality when $a/\eta = b/(1-\eta)$.
    The relaxed optimum is therefore $a+b = \sqrt{\alpha}$, achieved at $(a,b) = (\eta\sqrt{\alpha}, (1-\eta)\sqrt{\alpha})$.

    \textbf{Regime 1 ($\eta \le \sqrt{\alpha}$):}
    The relaxed optimum satisfies all linear constraints: $a = \eta\sqrt{\alpha} \le \alpha$ is equivalent to $\eta \le \sqrt{\alpha}$, and $a \le \eta$ and $b \le 1-\eta$ hold since $\sqrt{\alpha} \le 1$.
    Thus $\beta^* = \sqrt{\alpha}$.

    \textbf{Regime 2 ($\sqrt{\alpha} < \eta \le \eta_{\mathrm{crit}}$):}
    The relaxed optimum has $a = \eta\sqrt{\alpha} > \alpha$, violating the constraint $a \le \alpha$.
    Note that $a \le \eta$ is not the binding constraint, since $\eta > \sqrt{\alpha} > \alpha$ (using $\alpha < 1$).
    The optimum of the original problem must therefore have $a = \alpha$, and we maximize $b$ subject to the remaining constraints.
    The ellipsoid gives $\frac{b^2}{1-\eta} \le \alpha - \frac{\alpha^2}{\eta} = \frac{\alpha(\eta - \alpha)}{\eta}$, which is positive since $\eta > \alpha$.
    The largest $b$ satisfying the ellipsoid is $b_{\mathrm{ell}} \coloneqq \sqrt{(1-\eta)\cdot\frac{\alpha(\eta - \alpha)}{\eta}}$.
    We also require $b \le 1-\eta$, so $b^* = \min(b_{\mathrm{ell}},\; 1-\eta)$.

    We determine when $b_{\mathrm{ell}} > 1-\eta$.
    \begin{align*}
        (1-\eta)\cdot\frac{\alpha(\eta - \alpha)}{\eta} &> (1-\eta)^2 \\
        \frac{\alpha(\eta - \alpha)}{\eta} &> 1 - \eta \\
        \alpha(\eta - \alpha) &> \eta(1 - \eta) \\
        \eta^2 + (\alpha - 1)\eta - \alpha^2 &> 0.
    \end{align*}
    The left-hand side is a quadratic in $\eta$ with unique positive root $\eta_{\mathrm{crit}} = \frac{(1-\alpha)+\sqrt{1-2\alpha+5\alpha^2}}{2}$.
    For $\eta \le \eta_{\mathrm{crit}}$, we have $b_{\mathrm{ell}} \le 1-\eta$, so $b^* = b_{\mathrm{ell}}$ and $\beta^* = \alpha + \sqrt{(1-\eta)\bigl(\alpha - \frac{\alpha^2}{\eta}\bigr)}$.

    \textbf{Regime 3 ($\eta > \eta_{\mathrm{crit}}$):}
    Now $b_{\mathrm{ell}} > 1-\eta$, so the constraint $b \le 1-\eta$ binds before the ellipsoid.
    Thus $b^* = 1-\eta$ and $\beta^* = \alpha + 1 - \eta$.
\end{proof}

			Note that $\sqrt{\alpha}$ is the list-decoding threshold given by the standard Johnson bound (without hints).
	We interpret the three regimes as follows.
	In Regime~1 the hint is outside of the standard Johnson radius and is not helpful.
		As soon as the hint is within the Johnson radius (Regime~2), the hint can be used to improve the list-decoding threshold.
	As the hint gets even closer to the received word (Regime~3), we simply get the triangle inequality.
		Regime~2 is the most interesting, and we study the improvement in this regime over the standard Johnson bound.

\begin{theorem}\label{thm:johnson-comparison}
    Fix $\alpha \in (0,1)$. For $\eta = \sqrt{\alpha} + \epsilon$ with sufficiently small $\epsilon > 0$,
    \[
        \beta^*(\alpha, \eta) = \sqrt{\alpha} - \frac{\epsilon^2}{2(1-\sqrt{\alpha})} + O_\alpha(\epsilon^3).
    \]
	    In particular, a hint with agreement exceeding the Johnson radius by $\epsilon$ improves the list-decoding threshold by an additive term $\Theta(\epsilon^2)$.
\end{theorem}

\begin{proof}
    For sufficiently small $\epsilon$, we are in the second regime of Theorem~\ref{thm:solve-single-opt}, so
    $$\beta^* = \alpha + \sqrt{(1-\eta)\alpha(\eta - \alpha)/\eta}.$$
    The result follows by Taylor-expanding around $\eta = \sqrt{\alpha}$.
\end{proof}

\begin{figure}[h]
	    \begin{center}
	        \includegraphics[width=0.5\textwidth]{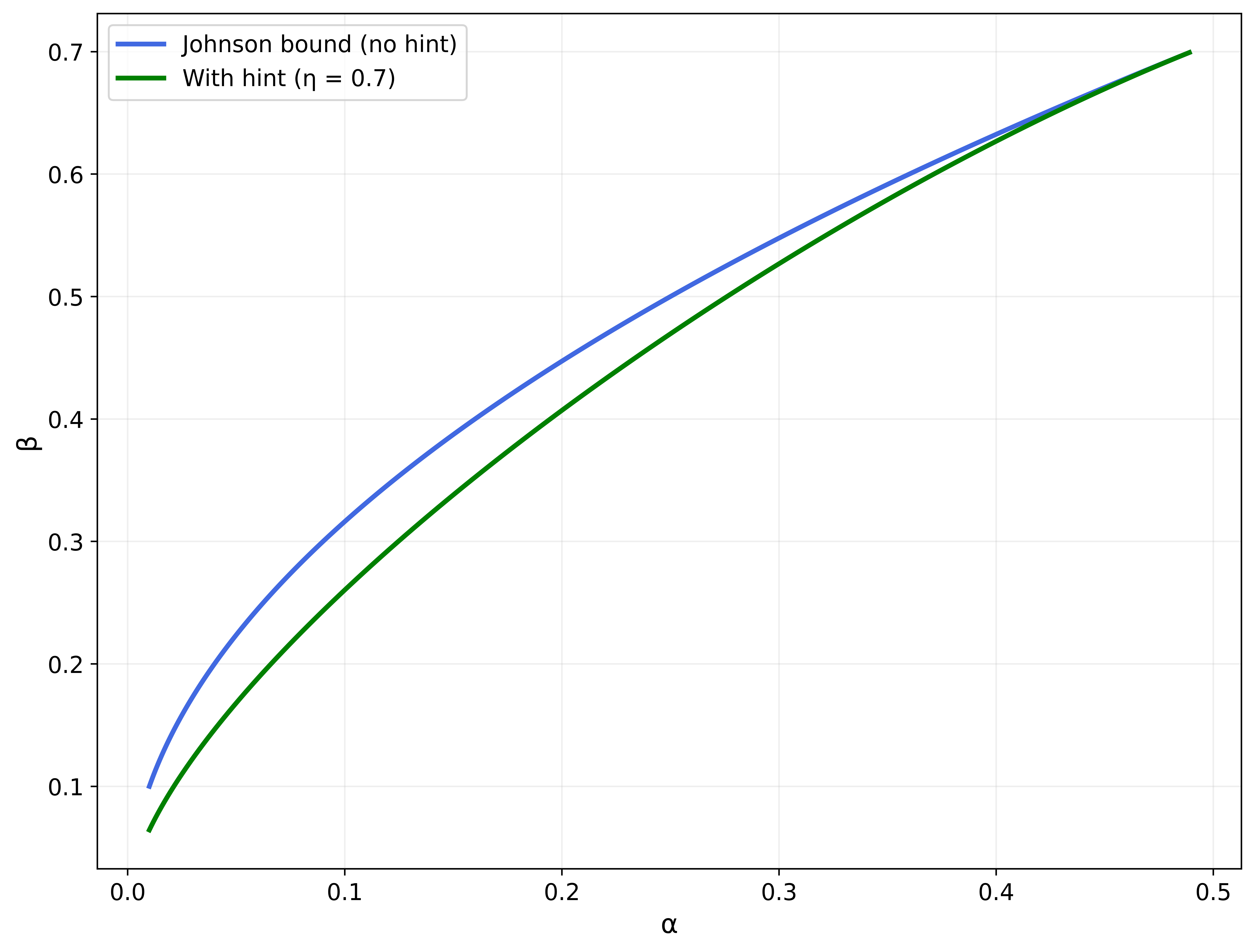}
	    \end{center}
	    \caption{List decoding with a hint of agreement $\eta = 0.7$.
	        The $x$-axis is $\alpha$ and the $y$-axis is the decoding threshold $\beta$.
	        The blue curve is the Johnson bound without a hint and the green curve is the Johnson bound with a hint of agreement $\eta$.}
	    \label{fig:single-hint}
	\end{figure}

Figure~\ref{fig:single-hint} illustrates the improvement.

Let $\beta_{\text{Johnson}} = \sqrt{\alpha}$ be the list-decoding threshold given by the Johnson bound.
As a corollary, we get a result on the structure of lists showing the tightness of the Johnson bound.
In particular, for any exponential-size list $L$ of codewords all with agreement at least $\beta_{\text{Johnson}} - \delta$,
all elements of $L$ have agreement at most $\beta_{\text{Johnson}} + O_\alpha(\sqrt{\delta})$.
Any element of the list with higher agreement could be used as a hint and, in conjunction with the previous theorems, would imply that the list size must be polynomial.

\begin{corollary}\label{cor:tight-johnson-uniform}
	    Let $C$ be a code with maximum agreement $\alpha$ over a large alphabet and let $r \in \Sigma^n$.
    For any sufficiently small constant $\delta > 0$, if $L \subset C$ is a list of $\exp(\Omega(n))$ codewords, each with agreement at least $\sqrt{\alpha} - \delta$ with $r$, then every codeword $c \in L$ satisfies $\agree(c, r) \le \sqrt{\alpha} + O_\alpha(\sqrt{\delta})$.
\end{corollary}

\begin{proof}
    Suppose some $c \in L$ has $\agree(c, r) = \sqrt{\alpha} + \epsilon$ for $\epsilon > 0$.
    Using $c$ as a hint, Theorem~\ref{thm:johnson-comparison} gives decoding threshold $\beta^* = \sqrt{\alpha} - \frac{\epsilon^2}{2(1-\sqrt{\alpha})} + O_\alpha(\epsilon^3)$.
    For $\epsilon \ge C\sqrt{\delta}$ with $C$ a sufficiently large constant depending on $\alpha$, we get $\beta^* < \sqrt{\alpha} - \delta$.
    The remaining $\exp(\Omega(n)) - 1$ codewords each have agreement at least $\sqrt{\alpha} - \delta > \beta^*$, so by Theorem~\ref{thm:large-alphabet-johnson-single}, the list has polynomial size, a contradiction.
\end{proof}

\begin{remark}
    For comparison, the triangle inequality gives only the weaker bound $\agree(c, r) \le \sqrt{\alpha} + (1 - \sqrt{\alpha})^2 + \delta$.
    Indeed, if $c_1, c_2 \in C$ have agreements $\beta_1, \beta_2$ with $r$, the positions where both agree with $r$ overlap in at least $(\beta_1 + \beta_2 - 1)n$ coordinates, so $\agree(c_1, c_2) \ge \beta_1 + \beta_2 - 1$.
    Since $\agree(c_1, c_2) \le \alpha$, we get $\beta_1 + \beta_2 \le 1 + \alpha$.
    Taking $\beta_2 \ge \sqrt{\alpha} - \delta$ yields $\beta_1 \le 1 + \alpha - \sqrt{\alpha} + \delta = \sqrt{\alpha} + (1 - \sqrt{\alpha})^2 + \delta$.
    Unlike Corollary~\ref{cor:tight-johnson-uniform}, this allows a constant gap of $(1 - \sqrt{\alpha})^2$ above $\sqrt{\alpha}$ even as $\delta \to 0$.
\end{remark}

\section{Johnson Bound with a hint, \texorpdfstring{$q$ small}{q small}} \label{sec:small-q}

The proof of the lower bound breaks down when $q$ is small because the agreement between codewords in the list on symbols that are not equal to $r$ and $h$ is no longer negligible.
As an extreme example, consider the case of $q = 2$.
Take $r$ to be the all-zeros string, and let $h$ be the string consisting of $\eta n$ 0s followed by $(1 - \eta) n$ 1s.
Let $a$ be the normalized agreement with $r$ on the first $\eta n$ coordinates, i.e., $a = \frac{1}{n}\sum_{i=1}^{\eta n} p_{i, 0}$.
Since $q = 2$, the remaining mass on these coordinates must all go to symbol~$1$, contributing $\eta - a$ in total.
By Cauchy-Schwarz, the pairwise agreement between codewords on these coordinates is at least $\frac{a^2}{\eta} + \frac{(\eta - a)^2}{\eta}$, whereas for large~$q$ the second term is $O(1/q)$ and can be absorbed into the slack.
Similarly, on the last $(1-\eta)n$ coordinates, the agreement with $r$ (symbol~$0$) and with $h$ (symbol~$1$) together determine the full distribution, and both contribute non-negligibly to pairwise agreement.

To handle the small-alphabet case, we adapt the proof by partitioning all of the $p_{i, \sigma}$ (not just those with $\sigma = r_i$).
In particular, we partition $[n] \times \Sigma$ into five parts as follows.

\begin{itemize}
    \item $U$: coordinates where $r_i = h_i$ and symbol $\sigma = r_i$.
    \item $V$: coordinates where $r_i = h_i$ and symbol $\sigma \neq r_i$.
    \item $W$: coordinates where $r_i \neq h_i$ and symbol $\sigma = r_i$.
    \item $Z$: coordinates where $r_i \neq h_i$ and symbol $\sigma = h_i$.
	    \item $Y$: coordinates where $r_i \neq h_i$ and symbol $\sigma \neq r_i$ and $\sigma \neq h_i$.
\end{itemize}

We define macroscopic variables $u, w, z$ as the normalized sums over parts $U, W$, and $Z$:
\[
    u \coloneqq \frac{1}{n} \sum_{(i,\sigma) \in U} p_{i,\sigma}, \qquad
    w \coloneqq \frac{1}{n} \sum_{(i,\sigma) \in W} p_{i,\sigma}, \qquad
    z \coloneqq \frac{1}{n} \sum_{(i,\sigma) \in Z} p_{i,\sigma}.
\]
Then the normalized sums over the remaining parts $V$ and $Y$ are $\eta - u$ and $1 - \eta - w - z$, respectively.
Applying Cauchy-Schwarz to each of the five groups, the sum $\frac{1}{n}\sum_{i,\sigma} p_{i,\sigma}^2$ is lower bounded by the quadratic expression in $u, w, z$ appearing as the ellipsoid constraint below. Together with the linear constraints from the hint agreement and the box constraints, this yields the following optimization for the exact list-decoding threshold.

\begin{optimization}\label{opt:johnson-small-alphabet-single-hint}
    \begin{align*}
        \max \quad        & u + w \\
        \text{s.t.} \quad & \frac{u^2 + \frac{(\eta - u)^2}{q-1}}{\eta}  + \frac{w^2 + z^2 + \frac{(1 - \eta - w - z)^2}{q-2}}{1 - \eta} \le \alpha \\
                          & u + z \leq \alpha                                                                                            \\
                          & u \leq \eta                                                                                                  \\
                          & w + z \leq 1 - \eta                                                                                          \\
                          & u, w, z \in [0,1]
    \end{align*}
\end{optimization}

Note the following convention for $q=2$.
On a coordinate where $r_i \neq h_i$, the two available symbols are precisely $r_i$ and $h_i$.
Accordingly, in the binary case $w+z=1-\eta$, and we omit the term
involving $q-2$ in the quadratic constraint, which avoids dividing by zero.

For parameters $(\alpha, \eta, q)$, let $\beta^*(\alpha, \eta, q)$ be the optimum of Optimization~\ref{opt:johnson-small-alphabet-single-hint}.

\begin{theorem}\label{thm:small-alphabet-johnson-single}
    Consider the list-decoding problem with a single hint over an alphabet of size $q\geq 2$. Let $\eta \in [0,1]$ be the agreement between the hint and the received word, and let $\alpha \in (0,1)$ be the maximum agreement of the code, satisfying $\alpha > 1/q$.
    Let $\beta^*(\alpha, \eta, q)$ be the optimum of Optimization~\ref{opt:johnson-small-alphabet-single-hint}.
	    Then $\beta^*(\alpha, \eta, q)$ is the average list-decoding threshold for the one-hint, small-alphabet list-decoding problem.
\end{theorem}

We omit the proof since it is almost identical to that of Theorem~\ref{thm:large-alphabet-johnson-single} and the key difference is described above.
The other difference is in the Slater point used. Instead of taking the all-zeros point, we take a point corresponding to the uniform distribution on the alphabet. In particular, we use the Slater point $(\eta/q, (1 - \eta)/q, (1 - \eta)/q)$.
This idea is also present in the most general case, Theorem~\ref{thm:large-alphabet-johnson-multi}, for which we give a full proof.

\section{Johnson Bound with multiple hints, \texorpdfstring{$q$}{q} large} \label{sec:multi-large-q}

When there are multiple hints $h_1, \ldots, h_m$, the received word may agree with multiple hints simultaneously.
In the single-hint case, each coordinate belongs to one of two groups depending on whether the hint agrees with the received word. With $m$ hints, each coordinate $i$ is characterized by the subset $S \subseteq [m]$ of hints that agree with $r$ at that coordinate, yielding up to $2^m$ groups.

Formally, for each $S \subseteq [m]$, define $U_S = \{i \in [n]: \{j: h_{j,i} = r_i\} = S\}$, and let $\eta_S = |U_S|/n$.
The vector $\eta \in [0,1]^{\mathcal{P}([m])}$ with $\sum_{S \subseteq [m]} \eta_S = 1$ records which hints agree with the received word.

The macroscopic variables and constraints are defined as before, using this finer partition of coordinates.
This yields the following optimization problem.

\begin{optimization}\label{opt:multi-hint-large-alphabet}
    The parameters are $\alpha \in [0,1]$, $m \in \N$, and $\eta \in [0,1]^{\mathcal{P}([m])}$.
    The decision variables are $(a_{S})_{S \subseteq [m]}$.
    \begin{align*}
        \max \quad        & \sum_{S \subseteq [m]} a_{S} \\
        \text{s.t.} \quad & \sum_{S \subseteq [m]} \frac{a_{S}^2}{\eta_{S}} \le \alpha \\
                          & \sum_{S \ni j} a_{S} \leq \alpha              & \forall j \in [m]     \\
                          & 0 \leq a_{S} \leq \eta_S                      & \forall S \subseteq [m] \\
    \end{align*}
\end{optimization}

Let $\beta^*(\alpha, m, \eta)$ be the optimum of Optimization~\ref{opt:multi-hint-large-alphabet}.

\begin{theorem}\label{thm:multi-hint-large-alphabet}
    Consider the list-decoding problem with $m$ hints over a large alphabet. Let $\eta \in [0,1]^{\mathcal{P}([m])}$ be an $\alpha$-realizable agreement profile between the hints and the received word, with $\eta_S$ denoting the fraction of coordinates where exactly the hints in $S$ agree with the received word, and let $\alpha \in (0,1)$ be the maximum agreement of the code.
    Let $\beta^*(\alpha, m, \eta)$ be the optimum of Optimization~\ref{opt:multi-hint-large-alphabet}.
	    Then $\beta^*(\alpha, m, \eta)$ is the average list-decoding threshold for the multiple-hint, large-alphabet list-decoding problem.
\end{theorem}

Again, the proof is almost identical to that of Theorem~\ref{thm:large-alphabet-johnson-single}, and the main difference is simply taking a finer partition of the coordinates based on which hints agree with the received word. For the lower bound, take $r$ to be constant and assign a distinct non-$r$ symbol to each hint wherever it disagrees with $r$. Then $\agree(h_j,h_k)=\sum_{S\supseteq\{j,k\}}\eta_S\leq\alpha$ by realizability. We omit the details.

\section{Johnson Bound with multiple hints, general case} \label{sec:general}

In the previous section, we partitioned the coordinates based on which hints agree with the received word. This suffices in the large-alphabet regime because on coordinates where a hint disagrees with $r$, the hint symbol is distinct from the received symbol and contributes negligibly to agreement.

When the alphabet is small, the situation is more subtle. Two hints that both disagree with $r$ at a given coordinate may place the \emph{same} non-received symbol there, creating additional structure. More generally, the agreement between codewords, the received word, and the hints depends not just on which hints match $r$, but on the full pattern of symbol equalities among all of them.

Consider $q = 3$ with $\Sigma = \{0, 1, 2\}$, and two hints $h_1, h_2$. Suppose $r_i = 0$, $h_{1,i} = 1$, and $h_{2,i} = 1$. Both hints disagree with $r$, so in the partition from the previous section, this coordinate belongs to the group $S = \emptyset$. However, the two hints place the \emph{same} symbol at this coordinate. A codeword placing symbol $1$ simultaneously agrees with both hints and disagrees with $r$, which is a stronger constraint than if the two hints placed distinct non-received symbols (say $h_{1,i} = 1$ and $h_{2,i} = 2$). The general case must distinguish these two configurations.

Thus, this most general case uses an even finer partition of coordinates based on the full pattern of symbol equalities among $r$ and all hints.
At each coordinate $i$, we record which of $r, h_1, \ldots, h_m$ place the same symbol, forming a partition of $\{0, 1, \ldots, m\}$ (where $0$ represents $r$).
Coordinates sharing the same partition type are grouped together.
We now set up the notation.

Let $r = h_0$ and $h_1, \ldots, h_m \in C$.
For each $i \in [n]$, let $\pi_i$ be the partition of $\{0,1,\ldots,m\}$ induced by the symbols appearing at coordinate $i$.
Formally, for each coordinate $i \in [n]$, define the equivalence relation $\sim_i$ on $\{0,1,\ldots,m\}$ such that $j \sim_i j'$ if and only if $h_{j,i} = h_{j',i}$.
$\pi_i$ is then the partition on $\{0,1,\ldots,m\}$ induced by this equivalence relation.
For each $i \in [n]$ and each part $S \in \pi_i$, define the symbol $\sigma_{i,S}$ to be the symbol that the members of $S$ take at coordinate $i$.
Let $\Pi$ be the set of all partitions of $\{0,1,\ldots,m\}$.
For each $\pi \in \Pi$, let $U_\pi = \{i \in [n] : \pi_i = \pi\}$.
The parameter $\eta \in [0,1]^{\Pi}$ records the frequency of each partition type: $\eta_\pi = |U_\pi| / n$ is the fraction of coordinates $i \in [n]$ with $\pi_i = \pi$.
We write $s_\pi$ for the number of parts in $\pi$.

\begin{optimization}[Multiple hints]\label{opt:multi}
    The parameters are $\alpha \in [0,1]$, an alphabet $\Sigma$ of size $q$, $m \in \N$, and $\eta \in [0,1]^{\Pi}$.
	    The decision variables are $(a_{\pi,S})_{\pi \in \Pi, S \in \pi}$ and $(b_\pi)_{\pi\in\Pi}$.
    \begin{align*}
        \max \quad        & \sum_{\substack{\pi \in \Pi                                                        \\ S \in \pi \\ S \ni 0}} a_{\pi, S} \\
        \text{s.t.} \quad & \sum_{\substack{\pi \in \Pi                                                        \\ S \in \pi}} \frac{a_{\pi, S}^2}{\eta_{\pi}} + \sum_{\substack{\pi \in \Pi \\ q > s_\pi}} \frac{b_\pi^2}{(q-s_\pi)\eta_\pi} \leq \alpha \\
                          & \sum_{\substack{\pi \in \Pi,\ S \in \pi                                            \\ S \ni j}} a_{\pi, S} \leq \alpha  & \forall j \in [m]\\
                          & b_\pi + \sum_{S \in \pi} a_{\pi, S} = \eta_{\pi}, & \forall \pi \in \Pi            \\
                          & a_{\pi, S}, b_{\pi} \in [0,1]                     & \forall \pi \in \Pi, S \in \pi
    \end{align*}
\end{optimization}

Let $\beta^*(\alpha, q, m, \eta)$ denote the optimum of Optimization~\ref{opt:multi}.

Before proving that $\beta^*$ is a list-decoding threshold, we record some facts about Optimization~\ref{opt:multi}. The first lemma establishes a Slater point, a strictly feasible solution, which guarantees strong duality and shows $\beta^* > 1/q$. The Slater point is based on the uniform distribution. The second lemma is a lower bound on the directional derivative at the optimal solution in the direction of a point $x$ that obtains a higher objective value and satisfies the linear constraints but not the ellipsoid constraint. This will be a key ingredient for the upper bound on list size and the decoding algorithm analysis.
Note that the assumption $\alpha > 1/q$ below is natural: by the Plotkin bound, any code over a $q$-ary alphabet with maximum agreement at most $1/q$ has only polynomially many codewords, so the list-decoding problem is only interesting when $\alpha > 1/q$.

\begin{lemma}[Slater point for Optimization~\ref{opt:multi}]\label{lem:slater-point}
    Suppose $\alpha > 1/q$.
    Define $z$ by $a_{\pi,S} = \eta_\pi/q$ for all $\pi \in \Pi, S \in \pi$ and $b_\pi = (q - s_\pi)\eta_\pi/q$ for all $\pi \in \Pi$.
    Then $z$ is feasible for Optimization~\ref{opt:multi} with $f(z) = 1/q < \alpha$ and objective $c^T z = 1/q$.
    Furthermore, $\beta^*(\alpha, q, m, \eta) > 1/q$.
\end{lemma}

\begin{proof}
    The equality constraints hold: $\sum_{S \in \pi} a_{\pi,S} + b_\pi = s_\pi \eta_\pi / q + (q - s_\pi)\eta_\pi / q = \eta_\pi$.
	    The capacity constraints give $\sum_{\pi,\, S \ni j} a_{\pi,S} = \sum_\pi \eta_\pi / q = 1/q \le \alpha$, and the box constraints are easily verified.
    The ellipsoid value is
    \[
        f(z) \;=\; \sum_{\pi, S} \frac{\eta_\pi}{q^2} + \sum_{\pi:\, q > s_\pi} \frac{(q - s_\pi)\eta_\pi}{q^2}
        \;=\; \sum_\pi \frac{s_\pi \eta_\pi + (q - s_\pi)\eta_\pi}{q^2}
        \;=\; \frac{1}{q}
        \;<\; \alpha.
    \]
    Since $f(z) < \alpha$, we can increase the objective beyond $1/q$ while remaining feasible.
    Pick any $\pi$ with $\eta_\pi > 0$ and any $S \in \pi$ with $0 \in S$.
    If $q > s_\pi$ (so that $b_\pi = (q - s_\pi)\eta_\pi/q > 0$), increase $a_{\pi,S}$ by a small $\mu > 0$ and decrease $b_\pi$ by $\mu$.
    If $q = s_\pi$ (so that $b_\pi = 0$), pick any $S' \in \pi$ with $0 \notin S'$ (which exists since $s_\pi = q \ge 2$), and increase $a_{\pi,S}$ by $\mu$ while decreasing $a_{\pi,S'}$ by $\mu$.
    In both cases, the equality constraint is preserved and the objective increases by $\mu$.
    The capacity constraints remain satisfied for small $\mu$, since at $z$ each has value $1/q < \alpha$.
    The ellipsoid and box constraints are likewise preserved for $\mu$ small enough.
    Thus $\beta^* > c^T z = 1/q$.
\end{proof}

\begin{lemma}\label{lem:generalized-first-order}
    Consider the optimization problem
    \[
        \max_{x \in \mathcal{X}} \; c^T x \quad \text{subject to} \quad f(x) \le \alpha,
    \]
    where $f$ is convex and differentiable and $\mathcal{X}$ is a convex set defined by linear constraints.
    Let $x^*$ be an optimal solution with value $\beta^* = c^T x^*$.
    If there exists a Slater point $z \in \mathcal{X}$ with $f(z) < \alpha$, and a point $x \in \mathcal{X}$ with $c^T(x - x^*) = \epsilon > 0$, then
    \[
        \langle \nabla f(x^*),\, x - x^* \rangle \;\ge\; \frac{\epsilon\,(\alpha - f(z))}{\beta^* - c^T z}.
    \]
    Furthermore, $f(x^*) = \alpha$.
    \end{lemma}

\begin{proof}
    Since $z \in \mathcal{X}$ satisfies $f(z) < \alpha$, Slater's condition holds for the convex constraint $f(x) \le \alpha$.
    Combined with the linearity of the constraints defining $\mathcal{X}$, strong duality applies and there exists a KKT multiplier $\lambda \ge 0$ such that $x^*$ maximizes the Lagrangian
    \[
        L(x) \;=\; c^T x - \lambda\bigl(f(x) - \alpha\bigr)
    \]
    over $x \in \mathcal{X}$, with complementary slackness $\lambda(f(x^*) - \alpha) = 0$.

    We first show $\lambda > 0$.
    If $\lambda = 0$, then $x^*$ maximizes $c^T x$ over $\mathcal{X}$ unconditionally, so $c^T x \le \beta^*$ for all $x \in \mathcal{X}$, contradicting $c^T(x - x^*) = \epsilon > 0$.
    So $\lambda > 0$.
    Then, by complementary slackness, we have $f(x^*) = \alpha$, establishing the ``furthermore'' part of the lemma.

    Since $x^*$ maximizes $L$ over the convex set $\mathcal{X}$, the first-order necessary condition for optimality gives $\langle \nabla L(x^*),\, x - x^* \rangle \le 0$.
    Expanding $\nabla L(x^*) = c - \lambda \nabla f(x^*)$ yields
    \[
        \epsilon \;=\; c^T(x - x^*) \;\le\; \lambda\,\langle \nabla f(x^*),\, x - x^* \rangle.
    \]

    We now upper bound $\lambda$.
    Since $x^*$ maximizes $L$ over $\mathcal{X}$ and $z \in \mathcal{X}$, we have $L(z) \le L(x^*)$.
    Using $f(x^*) = \alpha$, this gives
    \[
        c^T z - \lambda\bigl(f(z) - \alpha\bigr) \;\le\; \beta^*,
    \]
    so $\lambda(\alpha - f(z)) \le \beta^* - c^T z$, and therefore
    \[
        \lambda \;\le\; \frac{\beta^* - c^T z}{\alpha - f(z)}.
    \]

	    Combining these inequalities, we obtain
    \[
        \langle \nabla f(x^*),\, x - x^* \rangle
        \;\ge\; \frac{\epsilon}{\lambda}
        \;\ge\; \frac{\epsilon\,(\alpha - f(z))}{\beta^* - c^T z}. \qedhere
    \]
\end{proof}

\begin{theorem}[Johnson bound with multiple hints] \label{thm:large-alphabet-johnson-multi}
    Consider the list-decoding problem with $m$ hints over an alphabet of size $q$. Let $\eta \in [0,1]^{\Pi}$ be a $(q,\alpha)$-realizable partition profile between the received word and the hints, and let $\alpha \in (0,1)$ be the maximum agreement of the code.
    Suppose $\alpha > 1/q$.
    Let $\beta^*(\alpha, q, m, \eta)$ be the optimum of Optimization~\ref{opt:multi}.
	    Then $\beta^*(\alpha, q, m, \eta)$ is the average list-decoding threshold for the list-decoding problem with parameters $(\alpha, q, m, \eta)$.
\end{theorem}

\begin{proof}[Proof of the upper bound ($\beta > \beta^*$)]
	    Let $(C, r, h_1, \ldots, h_m)$ be an instance of the list-decoding problem with multiple hints and parameters $(\alpha, q, m, \eta)$.
    Let $L \subset C \setminus \{h_1, \ldots, h_m\}$ be such that $\E_{c \sim L}[\agree(c, r)] \geq \beta$ where $\beta = \beta^* + \epsilon$ for $\epsilon > 0$.

    For $i\in[n]$ and $\sigma\in\Sigma$, set $p_{i,\sigma} \coloneqq \Pr_{x\sim L}[x_i=\sigma]$.
    By Lemma~\ref{lem:expected-agreement-distinct},
    \[
        \frac{1}{n}\sum_{i\in[n]}\sum_{\sigma\in\Sigma} p_{i,\sigma}^2 \;\le\; \alpha + \frac{1}{|L|}.
    \]

    Define
    \[
        a_{\pi,S} \;\coloneqq\; \frac{1}{n}\sum_{i \in U_\pi} p_{i,\sigma_{i,S}}, \qquad
        b_\pi \;\coloneqq\; \frac{1}{n}\sum_{i \in U_\pi} \sum_{\substack{\sigma \in \Sigma \\ \sigma \neq \sigma_{i,S}\,\forall S}} p_{i,\sigma}.
    \]
    Note that $(a, b)$ defined in this way satisfy all of the linear and box constraints.
    The equality constraints hold since $\sum_{S \in \pi} a_{\pi,S} + b_\pi = \frac{1}{n}\sum_{i \in U_\pi} \sum_{\sigma} p_{i,\sigma} = \frac{|U_\pi|}{n} = \eta_\pi$.
    The objective is $\sum_{\pi,\, S \ni 0} a_{\pi,S} = \frac{1}{n}\sum_i p_{i,r_i} = \E_{c \sim L}[\agree(c,r)] \ge \beta$.
    The capacity constraints hold: for each $j \in [m]$, $\sum_{\pi,\, S \ni j} a_{\pi,S} = \frac{1}{n}\sum_i p_{i,h_{j,i}} \le \alpha$ since every $c \in L \subset C$ satisfies $\agree(c, h_j) \le \alpha$.
    For the ellipsoid constraint, Cauchy-Schwarz applied to each group gives
    \[
        \frac{a_{\pi,S}^2}{\eta_\pi} \;\le\; \frac{1}{n}\sum_{i \in U_\pi} p_{i,\sigma_{i,S}}^2, \qquad
        \frac{b_\pi^2}{(q - s_\pi)\eta_\pi} \;\le\; \frac{1}{n}\sum_{i \in U_\pi} \sum_{\substack{\sigma \neq \sigma_{i,S}\,\forall S}} p_{i,\sigma}^2.
    \]

    Summing over all $\pi$ and $S$ gives
    \begin{equation} \label{eq:cs-general}
        f(a,b) \;\coloneqq\; \sum_{\pi \in \Pi} \sum_{S \in \pi} \frac{a_{\pi,S}^2}{\eta_\pi} + \sum_{\substack{\pi \in \Pi \\ q > s_\pi}} \frac{b_\pi^2}{(q-s_\pi)\eta_\pi} \;\le\; \frac{1}{n}\sum_{i \in [n]} \sum_{\sigma \in \Sigma} p_{i,\sigma}^2 \;\le\; \alpha + \frac{1}{|L|}.
    \end{equation}

    Let $x = (a, b)$, and $x^* = (a^*, b^*)$.
    To establish an upper bound of $\frac{1}{\epsilon(\alpha - 1/q)}$ on $|L|$, it suffices to show that $f(x) \geq \alpha + \epsilon(\alpha - 1/q)$.
	    The argument proceeds similarly to that in the single-hint case.

    By the first-order condition for convexity, $f(x) \ge f(x^*) + \langle \nabla f(x^*),\, x - x^* \rangle$.
    Let $\mathcal{X}$ denote the polytope defined by the linear, equality, and box constraints of Optimization~\ref{opt:multi}.
    Note that the objective is linear, so we can write it as $c^Tx$ for some $c$.
    By Lemma~\ref{lem:generalized-first-order} with the Slater point $z$ from Lemma~\ref{lem:slater-point} (satisfying $f(z) = 1/q$ and $c^T z = 1/q$) and $x \in \mathcal{X}$ with $c^T(x - x^*) \ge \epsilon$, we have
    \[
        \langle \nabla f(x^*),\, x - x^* \rangle \;\ge\; \frac{\epsilon\,(\alpha - 1/q)}{\beta^* - 1/q}, \text{ and } \quad  f(x^*) = \alpha.
    \]

    Hence, $f(a, b) \ge \alpha + \frac{\epsilon\,(\alpha - 1/q)}{\beta^* - 1/q} \geq \alpha + \epsilon(\alpha - 1/q)$, as required.

\end{proof}

\begin{remark}[List size]
    The proof gives the quantitative bound
    \[
        |L| \le \frac{\beta^* - 1/q}{(\alpha - 1/q)(\beta-\beta^*)}.
    \]
\end{remark}

\begin{proof}[Proof of the lower bound]
    Fix $\epsilon > 0$ and suppose $\beta = \beta^* - \epsilon$.
    It suffices to consider $\beta>1/q$. Indeed, if $\beta\leq 1/q$, choose any $\widetilde\beta\in(1/q,\beta^*)$ and apply the construction below with $\widetilde\beta$ in place of $\beta$; since every resulting codeword has agreement at least $\widetilde\beta>\beta$, the same instance witnesses the claimed lower bound at $\beta$. We may therefore assume $\beta>1/q$, so $0<\epsilon<\beta^*-1/q$.
    Let $(a^*, b^*)$ be an optimal solution to Optimization~\ref{opt:multi} with $f(a^*, b^*) \le \alpha$.
    Let $z$ be the Slater point from Lemma~\ref{lem:slater-point}, with $f(z) = 1/q < \alpha$, objective $c^T z = 1/q$.
    We define a new feasible point $(a, b)$ with objective $\beta^* - \epsilon$ by taking a convex combination of $(a^*, b^*)$ and $z$.
    Namely, let $t \coloneqq \epsilon / (\beta^* - 1/q)$ and define $(a, b) \coloneqq (1-t)(a^*, b^*) + t\,z$.
    Since the equality constraints are affine, $(a, b)$ is feasible with objective $(1-t)\beta^* + t/q = \beta^* - \epsilon = \beta$.
    For the ellipsoid, convexity of $f$ gives
    \[
        f(a, b) \le (1-t)\,f(a^*, b^*) + t\,f(z)
        \le (1-t)\alpha + \frac{t}{q}
        = \alpha - t\!\left(\alpha - \frac{1}{q}\right)
        = \alpha - \delta,
    \]
    where $\delta \coloneqq \frac{\epsilon\,(\alpha - 1/q)}{\beta^* - 1/q}  = O(\epsilon) > 0$.

    Let $r,h_1,\ldots,h_m$ be a witness to the $(q,\alpha)$-realizability of $\eta$. Since only the coordinatewise equality patterns matter, choose the witness in canonical form so that, for each $\pi\in\Pi$ and $S\in\pi$, the symbol $\sigma_{i,S}$ is independent of $i\in U_\pi$; denote it by $\sigma_{\pi,S}$.
    We will construct a code $L$ of size $\exp(\Omega(n))$ such that the average agreement between $r$ and $L$ is at least $\beta$, and $L \cup \{h_1, \ldots, h_m\}$ is a code with maximum agreement at most $\alpha$.
    Assume $|U_\pi| = \Theta(n)$ for all $\pi \in \Pi$; otherwise, we can ignore the coordinates in blocks of size $o(n)$ since they contribute only $o(1)$ to the normalized agreements.

    Fix $\pi \in \Pi$.
    Define a type $t_\pi$ over $\Sigma$ by
    \[
        t_\pi(\sigma) \coloneqq
        \begin{cases}
            a_{\pi,S}n               & \text{if } \sigma = \sigma_{\pi,S} \text{ for some } S \in \pi, \\[6pt]
            \frac{b_\pi}{q - s_\pi}n & \text{otherwise.}
        \end{cases}
    \]

    The constraint $b_\pi+\sum_{S\in\pi}a_{\pi,S}=\eta_\pi$ ensures $\sum_{\sigma\in\Sigma} t_\pi(\sigma)=|U_\pi|$. (If $q=s_\pi$, we must have $b_\pi=0$, and we omit all terms dividing by $q-s_\pi$.)
    By Lemma~\ref{lem:random-type}, there exists a code $L_\pi\subset \Sigma^{U_\pi}$ of size $\exp(\Omega(|U_\pi|))$, all of whose codewords have type $t_\pi$ and whose maximum agreement (normalized by $|U_\pi|$) is at most
    \[
        \sum_{\sigma\in\Sigma} \Big(\frac{t_\pi(\sigma)}{|U_\pi|}\Big)^2 + \delta / 2
        = \sum_{S\in\pi} \Big(\frac{a_{\pi,S}}{\eta_\pi}\Big)^2 \; + \; (q-s_\pi)\Big(\frac{b_\pi/(q-s_\pi)}{\eta_\pi}\Big)^2 \; + \; \delta / 2
        = \sum_{S\in\pi} \frac{a_{\pi,S}^2}{\eta_\pi^2} \; + \; \frac{b_\pi^2}{(q-s_\pi)\,\eta_\pi^2} \; + \; \delta / 2.
    \]

    Let $N \coloneqq \min_\pi |L_\pi| = \exp(\Omega(n))$, and fix an arbitrary enumeration of each $L_\pi$.
    Define $L \coloneqq \{(c_\pi^{(j)})_{\pi \in \Pi} : j \in [N]\}$, where $c_\pi^{(j)}$ is the $j$-th element of $L_\pi$.
    Since each $L_\pi$ has distinct elements, distinct codewords in $L$ differ on every block.

    To conclude the proof, we show that $L$ satisfies the required properties.
	    First, we show that distinct codewords in $L$ have agreement at most $\alpha$.
	    For distinct $c,c'\in L$, since they differ on every block, the agreement normalized by $n$ decomposes as
    \begin{align*}
        \agree(c,c') & \le \sum_{\pi\in\Pi} \eta_\pi\Big( \sum_{S\in\pi} \frac{a_{\pi,S}^2}{\eta_\pi^2} + \frac{b_\pi^2}{(q-s_\pi)\,\eta_\pi^2} + \delta/2 \Big)                             \\
                     & = \sum_{\pi\in\Pi} \sum_{S\in\pi} \frac{a_{\pi,S}^2}{\eta_\pi} \; + \; \sum_{\pi\in\Pi} \frac{b_\pi^2}{(q-s_\pi)\,\eta_\pi} \; + \; \delta/2 \\
                     & = f(a, b) + \delta/2
        \;\le\; (\alpha - \delta) + \delta/2 \;=\; \alpha - \delta/2.
    \end{align*}
	    Then, for any $j \in [m]$ and $c \in L$,
    \[
        \agree(c,h_j)
        = \frac{1}{n}\sum_{\pi\in\Pi} t_\pi(\sigma_{\pi,S(j)})
        = \sum_{\substack{\pi\in\Pi,\ S\in\pi\\ S\ni j}} a_{\pi,S}
        \;\le\; \alpha,
    \]
	    where $S(j)$ denotes the part of $\pi$ (a subset of $\{0,1,\ldots,m\}$) containing $j$. Likewise,
    \[
        \agree(c,r) = \sum_{\substack{\pi\in\Pi,\ S\in\pi\\ S\ni 0}} a_{\pi,S} \;\ge\; \beta.
    \]
\end{proof}

\section{Decoding RS codes with hints}\label{sec:decoding}

For Reed-Solomon codes, we can make our bounds algorithmic.
Namely, we can decode to the new list-decoding thresholds given in Theorems~\ref{thm:large-alphabet-johnson-single} and~\ref{thm:multi-hint-large-alphabet} (note that Reed-Solomon codes fall into the large-alphabet case).
As mentioned in the introduction, hints arise naturally as part of list decoding.
Any codeword found in a first pass by a list-decoding algorithm can then be interpreted as a hint.

By setting weights on the coordinates based on the optimal solutions to our optimization problems, we reduce the problem of decoding with hints to a single call to the weighted Guruswami-Sudan algorithm, the performance guarantee of which we state below.

\begin{theorem}[\cite{GuruswamiThesis}]\label{thm:weighted-gs}
	    Let $C$ be a Reed-Solomon code of length $n$ and maximum agreement $\alpha$.
	    Let $w \in \mathbb{R}_{\geq 0}^n$ be nonnegative, and let $r$ be a received word.
    Let $\epsilon > 0$.
	    Then there is an algorithm that runs in time $\poly(n, q, 1/\epsilon)$ and outputs all codewords $c$ such that
    \begin{equation}\label{eq:weighted-gs}
        \sum_{i: c_i = r_i} w_i >  \sqrt{\alpha n \sum_{i = 1}^n w_i^2} + \epsilon w_{\max}
    \end{equation}
\end{theorem}

Note that for Reed-Solomon codes, $q \geq n$, so we are in the large-alphabet regime.

\begin{theorem}[Hint-Aware Decoding]\label{thm:decoding-multi-hint}
    Let $C$ be a Reed-Solomon code of length $n$ and maximum agreement $\alpha$.
	    Let $(r, h_1, \ldots, h_m)$ be an instance of the list-decoding problem with multiple hints and parameters $\alpha$, $q$, $m$, and $\eta \in [0,1]^{\mathcal{P}([m])}$.
	    Suppose $\beta > \beta^*$, where $\beta^*$ is the optimum of Optimization~\ref{opt:multi-hint-large-alphabet}.
	    Then there exists an algorithm that finds all codewords with agreement at least $\beta$ with $r$ and runs in time $\poly(n, q, 1/\delta)$, where $\delta \coloneqq \beta - \beta^*$.
\end{theorem}

\begin{proof}
    Let $H_\beta\coloneqq\{h_j:\agree(h_j,r)\geq\beta\}$, and include these supplied hints in the output. It remains to recover every codeword $c\in C\setminus\{h_1,\ldots,h_m\}$ with $\agree(c,r)\geq\beta$; for each such $c$, we have $\agree(c,h_j)\leq\alpha$ for every $j\in[m]$. If $\beta^*$ is determined by the linear constraints alone (i.e., $\sum_S a_S \le \beta^*$ is forced by the capacity and box constraints without the ellipsoid being tight), then no such non-hint codeword can have agreement exceeding $\beta^*$, so the algorithm returns $H_\beta$.

    Otherwise, the optimum is strictly constrained by the ellipsoid, so the optimal solution $a^* = (a_S^*)_{S \subseteq [m]}$ satisfies $f(a^*) \coloneqq \sum_{S \subseteq [m]} \frac{(a_S^*)^2}{\eta_S} = \alpha$.

    For each $S \subseteq [m]$, define the coordinate block $A_S \coloneqq \{i \in [n] : \{j \in [m] : h_{j,i} = r_i\} = S\}$, so $|A_S| = \eta_S n$.
    Define the weights
    \[
        w_i \;\coloneqq\; \frac{a_S^*}{\eta_S}, \qquad i \in A_S.
    \]
    Run the weighted Guruswami-Sudan algorithm (Theorem~\ref{thm:weighted-gs}) with these weights.
    We claim that every codeword $c\in C\setminus\{h_1,\ldots,h_m\}$ with $\agree(c,r) \ge \beta$ satisfies the decoding condition~\eqref{eq:weighted-gs}.

    First, the sum of squared weights is
    \[
        \sum_{i=1}^n w_i^2
        \;=\; \sum_{S \subseteq [m]} \eta_S n \cdot \frac{(a_S^*)^2}{\eta_S^2}
        \;=\; \sum_{S \subseteq [m]} \frac{(a_S^*)^2 n}{\eta_S}
        \;=\; \alpha n,
    \]
    so the RHS of~\eqref{eq:weighted-gs} is $\alpha n$. For a codeword $c$, define the fractional agreements on each block by
    \[
        a_S \;\coloneqq\; \frac{1}{n} \Agree_S(c, r) = \frac{1}{n}\sum_{i \in A_S} \1_{c_i = r_i}.
    \]
    Then $\sum_S a_S = \agree(c,r) \ge \beta > \beta^*$.
    For each $j \in [m]$, on coordinates $i \in A_S$ with $j \in S$ we have $h_{j,i} = r_i$, so $c_i = r_i$ implies $c_i = h_{j,i}$. Thus $\sum_{S \ni j} a_S \le \agree(c, h_j) \le \alpha$.
    The box constraints $0 \le a_S \le \eta_S$ hold by definition.
    So $a$ satisfies all of the linear constraints of Optimization~\ref{opt:multi-hint-large-alphabet} with $c^T a \ge \beta > \beta^* = c^T a^*$.

    The weighted agreement of $c$ with $r$ is
    \[
        \sum_{i:\, c_i = r_i} w_i
        \;=\; n \sum_{S \subseteq [m]} a_S \cdot \frac{a_S^*}{\eta_S}
        \;=\; \frac{n}{2}\,\langle \nabla f(a^*),\, a \rangle
        \;=\; n\alpha + \frac{n}{2}\,\langle \nabla f(a^*),\, a - a^* \rangle,
    \]
    where we used the fact that $\nabla f(a^*) = (2a_S^*/\eta_S)_{S \subseteq [m]}$, and $\frac{1}{2}\langle \nabla f(a^*), a^* \rangle = f(a^*) = \alpha$.

    Let $\delta_0\coloneqq\beta-\beta^*>0$ be the decoding gap. Since $c^T(a-a^*)\geq\delta_0$, Lemma~\ref{lem:generalized-first-order}, applied with the Slater point $z=0$, gives
    \[
        \langle \nabla f(a^*),\,a-a^*\rangle
        \;\geq\;\frac{\delta_0\alpha}{\beta^*}.
    \]
    Therefore,
    \[
        \sum_{i:\,c_i=r_i}w_i
        \;\geq\;n\alpha+\frac{n\delta_0\alpha}{2\beta^*}.
    \]
    Set the interpolation accuracy to
    \[
        \epsilon_{\mathrm{alg}}\coloneqq\frac{n\delta_0\alpha}{4\beta^*}.
    \]
    Since $w_{\max}\leq1$, the weighted agreement exceeds $n\alpha+\epsilon_{\mathrm{alg}}w_{\max}$, as required by~\eqref{eq:weighted-gs}. The algorithm runs in time $\poly(n,q,1/\epsilon_{\mathrm{alg}})=\poly(n,q,1/\delta_0)=\poly(n,q,1/\delta)$.
    Finally, add $H_\beta$ to the output of the weighted Guruswami--Sudan algorithm and remove duplicates.
\end{proof}


\newpage

\printbibliography

\appendix

\section{Alternate Decoding via Factored Interpolation}\label{sec:alternate-decoding}

In this section, we present an alternate algorithm for list decoding Reed-Solomon codes with hints. While Section~\ref{sec:decoding} uses the weighted Guruswami-Sudan algorithm, this approach instead factors the hint polynomial directly into the interpolation polynomial.

\subsection{Decoding from one hint}\label{sec:alternate-single}

Recall the setup from Section~\ref{sec:decoding}: let $C$ be a Reed--Solomon code of length $n$ over $\mathbb{F}_q$ ($q \ge n$) with maximum agreement $\alpha$ (so the degree bound $k$ satisfies $k-1 = \alpha n$).
Given a received word $r = (r_1, \ldots, r_n)$ and a single hint $h \in C$ with $\agree(h, r) = \eta$, we wish to find all codewords with agreement at least $\beta$ with $r$.

We partition coordinates into
\[
    U = \{i \in [n] : h(x_i) = r_i\}, \qquad V = [n] \setminus U,
\]
so $|U| = \eta n$ and $|V| = (1-\eta)n$.
The main insight behind this version of the decoding algorithm is that the hint polynomial $h(X)$ can appear explicitly as a factor of the interpolation polynomial in the standard Guruswami--Sudan algorithm, providing free vanishing at coordinates in $U$.

\paragraph{Algorithm.}
Fix $c \in [0,1]$ to be optimized later. Let $M$ be a positive integer, and assume $cM \in \mathbb{Z}$.
\begin{enumerate}
    \item \textbf{Interpolate.} Find a nonzero bivariate polynomial $Q(X,Y)$ of $(1, k{-}1)$-weighted degree at most $D_Q$ vanishing with multiplicity $M$ at each $(x_i, r_i)$ for $i \in V$, and multiplicity $M(1-c)$ at each $(x_i, r_i)$ for $i \in U$.

    \item \textbf{Root-find.} Output all polynomials $f(X)$ of degree $< k$ with $(Y - f(X)) \mid Q(X,Y)$ and $\agree(f, r) \ge \beta$.
\end{enumerate}

\begin{theorem}\label{thm:alternate-single}
    Every codeword $f \ne h$ with $\agree(f,r) \ge \beta$ is output, provided
    \[
        \beta \;>\; c\alpha + \sqrt{\alpha\bigl((1-\eta) + (1-c)^2\eta\bigr)}.
    \]
\end{theorem}

\begin{proof}
    Define $P(X,Y) = (Y - h(X))^{cM} \cdot Q(X,Y)$.

    The proof is essentially identical to the proof of the original Guruswami--Sudan algorithm, with the only new ingredient being the different vanishing multiplicities on $U$ and $V$.

    \textbf{Multiplicity.}
	    For $i \in V$, the polynomial $Q$ vanishes to order $M$, so $P$ does as well.
	    For $i \in U$, we have $h(x_i) = r_i$, so $(Y - h(X))^{cM}$ vanishes to order $cM$ at $(x_i, r_i)$, and $Q$ vanishes to order $M(1-c)$. The product vanishes to order $cM + M(1-c) = M$.

    \textbf{Degree.}
    Since $h(X)$ has degree $\le k-1$, the polynomial $Y - h(X)$ has $(1,k{-}1)$-weighted degree $k-1$.
	    Thus, $P$ has weighted degree at most $cM(k-1) + D_Q$.

    For a codeword $f$ with $\agree(f,r) \ge \beta$, the univariate polynomial
    \[
        R(X) = (f(X) - h(X))^{cM} \cdot Q(X, f(X))
    \]
    has degree at most $cM(k-1) + D_Q$ and vanishes to order $\ge M$ at the $\ge \beta n$ points where $f(x_i) = r_i$.

    The interpolation cost for $Q$ is approximately $\frac{M^2 n}{2}((1-\eta) + (1-c)^2\eta)$, and the monomial count at weighted degree $D_Q$ is approximately $D_Q^2/(2(k-1))$.
    Setting these equal gives
    $$D_Q = M\sqrt{n(k-1)((1-\eta) + (1-c)^2\eta)}.$$
	    Thus, $R \equiv 0$ whenever $M\beta n > cM(k-1) + D_Q$, which simplifies to $\beta > c\alpha + \sqrt{\alpha((1-\eta) + (1-c)^2\eta)}$ (using $k - 1 = \alpha n$).

    Since $f \ne h$, the factor $(f(X) - h(X))^{cM}$ is a nonzero polynomial.
	    Thus, $R \equiv 0$ implies $Q(X, f(X)) \equiv 0$, i.e., $(Y - f(X)) \mid Q(X,Y)$.
\end{proof}

We now find the threshold $\beta^*_{\mathrm{fact}}$ by minimizing $c\alpha+\sqrt{\alpha((1-\eta)+(1-c)^2\eta)}$ over $c \in [0,1]$.

We have

\[
    \beta'(c) = \alpha - \frac{\sqrt{\alpha}\,\eta(1-c)}{\sqrt{(1-\eta)+(1-c)^2\eta}}.
\]
Since
\[
    \beta''(c) = \frac{\sqrt{\alpha}\,\eta(1-\eta)}{\bigl((1-\eta)+(1-c)^2\eta\bigr)^{3/2}} > 0,
\]
\(\beta(c)\) is strictly convex, so its minimum is either at \(c=0\) or at the unique interior critical point.

If \(\eta \le \sqrt{\alpha}\), then
\[
    \beta'(0) = \alpha - \eta\sqrt{\alpha} = \sqrt{\alpha}(\sqrt{\alpha}-\eta) \ge 0,
\]
so the minimum is attained at \(c=0\), giving \(\beta^*=\sqrt{\alpha}\).

If \(\eta > \sqrt{\alpha}\), then \(\beta'(0)<0\), so the minimum occurs at the unique critical point satisfying
\[
    (1-c)^2 = \frac{\alpha(1-\eta)}{\eta(\eta-\alpha)}.
\]
Substituting this value yields
\[
    \beta^* = \alpha + \sqrt{\frac{\alpha(1-\eta)(\eta-\alpha)}{\eta}}.
\]

This matches the first two regimes of Theorem~\ref{thm:solve-single-opt}. For the third regime, note that the triangle inequality implies that when $\beta > 1 - \eta + \alpha$, the only codeword with agreement $\geq \beta$ is $h$ itself. Hence, list-decoding to agreement at least $\beta$ is trivial. To recover the third regime, we simply compare the expression for $\beta^*$ in the $\eta > \sqrt{\alpha}$ regime with $1 - \eta + \alpha$, taking the smaller of the two. We find that $1 - \eta + \alpha$ is smaller so long as $\eta > \eta_{\text{crit}}$, where $\eta_{\mathrm{crit}}$ is as defined in Theorem~\ref{thm:solve-single-opt}.

\subsection{Generalization to Multiple Hints}\label{sec:alternate-multi}

It is natural to ask whether this factored interpolation approach generalizes to multiple hints. For two hints $h_1, h_2$, we could introduce multipliers $c_1, c_2 \ge 0$ (with $c_1 + c_2 \le 1$) and explicitly factor $(Y - h_1(X))^{c_1 M} (Y - h_2(X))^{c_2 M}$ out of the interpolant.
We partition the coordinates based on which hints agree with the received word $r$: let $\eta_\emptyset$ be the fraction where neither agrees, $\eta_1$ where only $h_1$ agrees, $\eta_2$ where only $h_2$ agrees, and $\eta_{12}$ where both agree.

The resulting threshold from optimizing the multipliers $c_1, c_2$ would be
\begin{align*}
    \beta^*_{\mathrm{fact}} = \min_{c_1, c_2} \quad & \alpha(c_1 + c_2) + \sqrt{\alpha \left[ \eta_\emptyset + \eta_1(1-c_1)^2 + \eta_2(1-c_2)^2 + \eta_{12}(1-c_1-c_2)^2 \right]} \\
    \text{subject to} \quad & c_1 \ge 0, \quad c_2 \ge 0, \\
    & c_1 + c_2 \le 1.
\end{align*}

This minimization is reminiscent of the Lagrangian dual of Optimization~\ref{opt:multi-hint-large-alphabet}: the variables $c_1$ and $c_2$ act like multipliers for the two capacity constraints. However, this optimization does not dualize the box constraints $a_S \le \eta_S$, which leads to a gap between $\beta^*_{\mathrm{fact}}$ and the original (primal) optimum, $\beta^*$.

As an example, consider an instance with two hints where $\alpha = 0.5$ and
\[
    \eta_\emptyset = 0, \qquad \eta_1 = 0.60, \qquad \eta_2 = 0.01, \qquad \eta_{12} = 0.39.
\]

Solving the primal formulation (maximizing over variables $a_S$) yields a primal optimum of $\beta^* = 0.51$, achieved at $a_1^* = 0.11, a_2^* = 0.01, a_{12}^* = 0.39, a_\emptyset^* = 0$. Note that in this solution, the box constraint for the second hint ($a_2 \le \eta_2$) is tight.
However, solving for $\beta_{\mathrm{fact}}^*$ yields a global optimum at $c_1^* \approx 0.8984$ and $c_2^* = 0$, giving $\beta^*_{\mathrm{fact}} \approx 0.5497$.
The gap arises because the factored formulation dualizes the capacity constraints (via multipliers $c_1, c_2$) but not the box constraints. More fundamentally, a global factor such as $Y - h_j(X)$ cannot isolate just the coordinates where hint $h_j$ is uniquely correct: any such factor acts on every coordinate where $h_j$ agrees, regardless of whether other hints also agree there. This lack of coordinate-level selectivity is exactly what the weighted Guruswami--Sudan algorithm provides, and it is why a purely factored approach cannot close the gap for multiple hints.

\end{document}